\documentclass[fleqn,aps,prb,twocolumn,showpacs,floatfix,longbibliography]{revtex4-2}
\usepackage{amsmath,amssymb,amsfonts,float,graphics,epsfig,epstopdf,color,verbatim,tabularx,bm,multirow,appendix,hyperref}

\usepackage{amsmath}
\usepackage{amssymb}
\usepackage{mathtools}
\usepackage{graphicx}
\usepackage{lipsum}
\usepackage{bm}
\usepackage{hyperref}
\usepackage{url}
\usepackage[utf8]{inputenc}
\usepackage{subfigure}
\usepackage{slashed,bbm}
\usepackage{graphics,psfrag,epsfig}
\usepackage{dsfont}
\usepackage{setspace}
\usepackage{wasysym}
\usepackage{slashed}
\usepackage{lipsum}
\usepackage{braket}
\usepackage{physics}
\usepackage{enumerate}%
\usepackage{makecell}
\usepackage{xcolor}

\definecolor{dark-red}{rgb}{0.4,0.15,0.15}
\definecolor{dark-blue}{rgb}{0.15,0.15,0.4}
\definecolor{medium-blue}{rgb}{0,0,0.5}
\hypersetup{
colorlinks, linkcolor={dark-blue},
citecolor={dark-blue}, urlcolor={medium-blue}
}
\usepackage{ulem}

\newcommand{\be}{\begin{equation}}
\newcommand{\ee}{\end{equation}}
\newcommand{\bea}{\begin{eqnarray}}
\newcommand{\eea}{\end{eqnarray}}

\renewcommand{\i}{\text{i}}

\begin{document}

\title{
Analogs of deconfined quantum criticality for non-invertible symmetry breaking in 1d
}

\author{Yu-Hsueh Chen}
\affiliation{Department of Physics, University of California at San Diego, La Jolla, California 92093, USA}
\author{Tarun Grover}
\affiliation{Department of Physics, University of California at San Diego, La Jolla, California 92093, USA}

\begin{abstract}
The spontaneous breaking of non-invertible symmetries  can lead to exotic phenomena such as coexistence of order and disorder. Here we explore second-order phase transitions in 1d spin chains between two phases that correspond to distinct patterns of non-invertible symmetry breaking. The critical point shares several features with well-understood examples of deconfined quantum critical points, such as enlarged symmetry and identical exponents for the two order parameters participating in the transition. Interestingly, such deconfined transitions involving non-invertible symmetries allow one to construct a \textit{whole family} of similar critical points by gauging spin-flip symmetries.  By employing gauging and bosonization, we characterize the phase diagram of our model in the vicinity of the critical point.  We also explore proximate phases and phase transitions in related models, including a deconfined quantum critical point between invertible order parameters that is enforced by a non-invertible symmetry.

\end{abstract}

\maketitle
\tableofcontents

\section{Introduction}

Non-invertible symmetries \cite{schafernameki2023ictp,shao2024whats} have attracted considerable attention as they introduce 
new anomalies and impose stringent constraints on phase diagrams 
\cite{bhardwaj2018finite,chang2019topological,tachikawa2020gauging,thorngren2024fusion,thorngren2024fusion2,zhang2023anomalies}.
A paradigmatic example is the Kramers--Wannier (KW) symmetry at the critical point of the Ising model \cite{kramers1941statistics}. Notably, the KW symmetry satisfies a Lieb--Schultz--Mattis type restriction \cite{chang2019topological,thorngren2024fusion,seiberg2024noninvertible}, implying any attempt to gap out the system while preserving KW symmetry must lead to degenerate ground states. One concrete example is the spontaneously breaking of the KW symmetry discussed in Ref.\cite{obrien2018lattice} {(see also Refs.\cite{grover2014emergent, rahmani2015emergent})}, which leads to the coexistence of order and disorder, a rather unusual phenomena \cite{bhardwaj2024categorical,bhardwaj2025gapped} that is otherwise ruled out~\cite{levin2020constraints}. In this paper, we explore new examples of phase transitions in which non-invertible symmetries play a crucial role and which admit a useful dual description. Notably, one of our examples involves breaking distinct non-invertible symmetries on either side of the critical point and has close analogies with deconfined quantum critical point (DQCP)~\cite{senthil2004deconfined, senthil2004quantum,senthil2023deconfine} -- the order parameters break symmetries unrelated to each other and the critical point has a self-duality symmetry that exchanges the order parameters. We also show that the DQCP involving non-invertible order parameters are robust under gauging spin-flip symmetries, which provides a machinery to generate a large family of new related DQCPs.

Recently, several works have explored the realization of lattice models with non-invertible symmetries, along with their associated phases and phase transitions~\cite{
obrien2020self, inamura2022on, eck2023critical, eck2024from, fechisin2023non, bhardwaj2023club,delcamp2024higher, bhardwaj2024illustrating, bhardwaj2024lattice, inamura20241+,seifnashri2024cluster, li2024noninvertible, chatterjee2024quantum, choi2025non, lu2024realizing,cao2024generating, pace2024lattice, warman2024categorical, meng2024non, gorantla2024tensor, antinucci2025symtft, cao2025global, gorantla2024duality, moy2024intertwined, bhardwaj2025gapped2,seifnashri2025gauging,su2025mathbb,lu2025strange,furukawa2025lattice,mana2025higher}. As hinted above, our primary motivation is to seek examples of phase transitions where the non-invertible symmetry is broken on either side of the transition, and which generalize the concept of DQCP beyond group-like symmetry. 
Let us therefore briefly review DQCP in the context of group-like symmetries.
DQCP describes a continuous single-parameter tuned transition between two phases that break different symmetries, a phenomenon that is generally prohibited within the conventional Landau--Ginzburg--Wilson framework. The key idea is that topological defects in one phase carry a fractional charge of the symmetry that is broken on the other side of the transition. Therefore, as topological defects proliferate, they destroy the order parameter on one side of the transition while simultaneously leading to non-zero order parameter on the other side.
Although DQCP was originally discovered in $(2+1)$-D systems, there is a growing interest in realizing analogous behavior in $(1+1)$-D models, which are  more tractable both analytically and numerically, thus providing a simpler perspective \cite{jiang2019ising, roberts2019deconfined, huang2019emergent,mudry2019quantum, roberts2021one,zhang2023exactly,su2023boundarycriticalitygaugingfinite,yang2025deconfined}. Notably, Ref.~\cite{zhang2023exactly} constructs an exactly solvable DQCP in $(1+1)$-D with non-onsite symmetries, realizable at the boundary of a $(2+1)$-D symmetry-protected topological state. This naturally motivates exploring the possibility of a DQCP at  the boundary of a $(2+1)$-D non-Abelian topologically ordered phase, which from the symmetry topological field theory perspective \cite{kong2020algebraic, ji2020categorical,gaiotto2021orbifold,apruzzi2023symmetry,freed2024topologicalsymmetryquantumfield} corresponds to a DQCP protected by non-invertible symmetries.
We remark that Ref.~\cite{chatterjee2024quantum} found numerical evidence for a transition between invertible and non-invertible SSB phases in $\text{Rep}(S_3)$ spin chains (see also related works Refs.~\cite{bhardwaj2024illustrating, bhardwaj2024lattice}). Here, we are primarily interested in DQCPs between two non-invertible SSB phases and on gaining an analytical understanding of such transitions.

\begin{figure}
\centering
\includegraphics[width=0.85\linewidth]{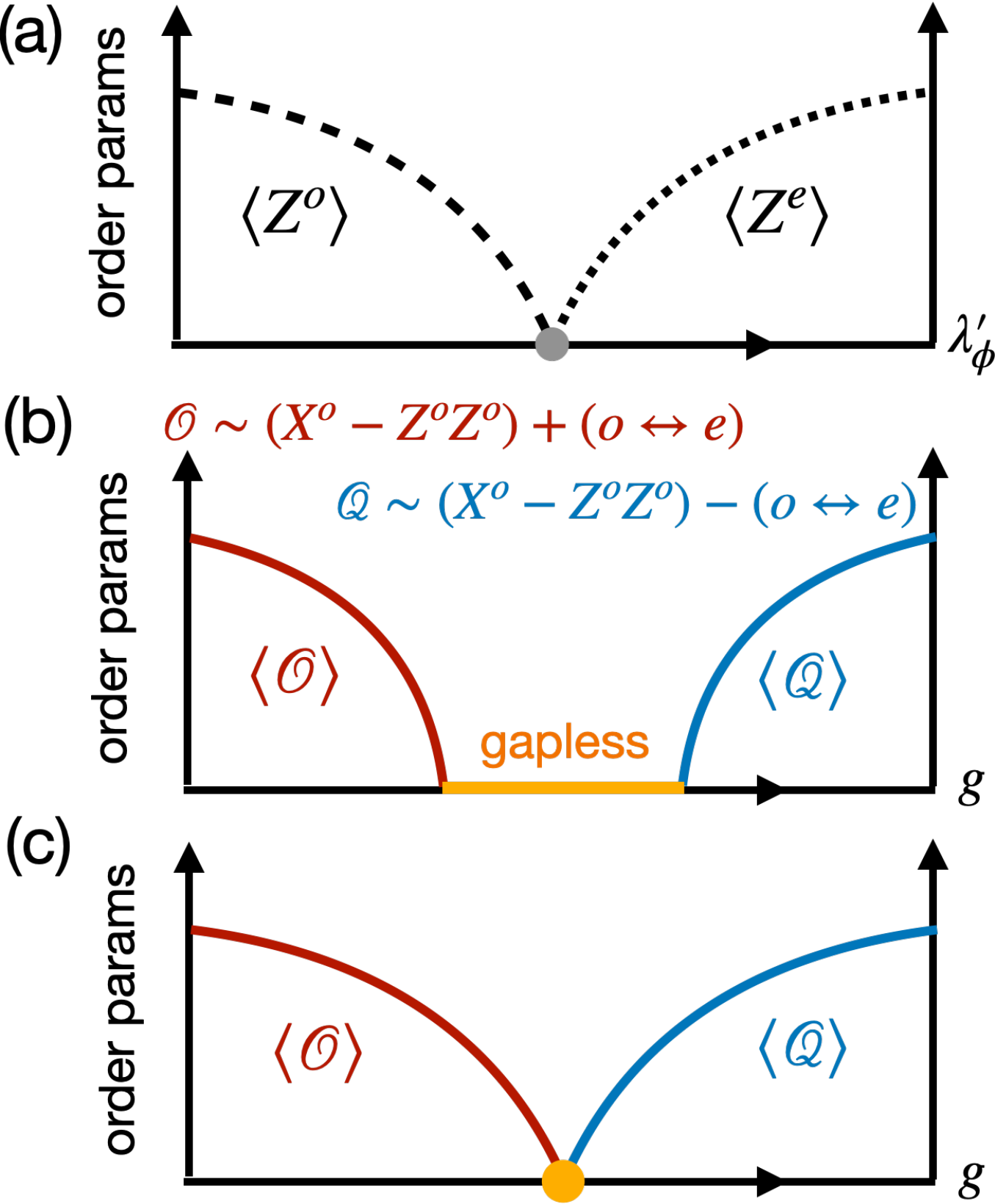}
\caption{Main results of the paper.
(a) Imposing $S D^o D^e$ as an additional symmetry generator on top of a $\mathbb{Z}^o_2 \times \mathbb{Z}^e_2$ symmetric spin chain induces a second-order transition tuned by a single parameter $\lambda_{\phi'}$ [see Eq.\eqref{Eq:singe_copy_Sint}] between two conventional symmetry-breaking phases.  
(b) Imposing $S D^o D^e$ and $S$ as additional symmetry generators leads to two distinct non-invertible SSB phases separated by a stable gapless regime.  
(c) By considering two coupled $\mathbb{Z}^o_2 \times \mathbb{Z}^e_2$ symmetric spin chains with similar non-invertible symmetries imposed in (b), a second-order transition tuned by a single parameter $g$ exists [see Eq.\eqref{Eq:rewrtie_Lint} and Fig.\ref{fig:DQCP}] between two non-invertible SSB phases.
}
\label{fig:main_results}
\end{figure}

In this paper, we investigate the possibility of continuous transition between two SSB phases in $\mathbb{Z}^o_2 \times \mathbb{Z}^e_2$ spin chains with additional symmetry generators that are non-invertible. Our main results are summarized in Fig.\ref{fig:main_results}. In the first scenario, we consider a $\mathbb{Z}^{o}_2 \times \mathbb{Z}^{e}_2$ symmetric spin chain where $o$ and $e$ denote the odd and even sites respectively. In addition, we  impose the non-invertible symmetry $S D^o D^e$, where $S$ denotes the swap between the even and the odd sites, while $D^o, D^e$ denote the KW symmetries acting on the odd and even sites respectively. We find a single-parameter tuned continuous transition between two conventional (i.e., group-like) symmetry-breaking phases [Fig.~\ref{fig:main_results}(a)]. This furnishes an example of DQCP that is enforced by a non-invertible symmetry. 
To explore transitions that involve spontaneous breaking of the non-invertible symmetry itself, we next further impose $S$ as an independent symmetry generator. We find two non-invertible SSB phases, which we refer to as $(D^o D^e, S D^oD^e)$-breaking and $(S, D^o D^e)$-breaking phases, characterized by distinct order parameters that are charged under their respective non-invertible symmetry. 
However, the two non-invertible SSB phases are separated by a stable, gapless regime, thus ruling out a generic direct transition between them [Fig.\ref{fig:main_results}(b)].
By considering \textit{two copies} of $\mathbb{Z}_2 \times \mathbb{Z}_2$ spin chains with similar non-invertible symmetries, we find that the interactions between the two copies can generate relevant perturbations that destabilize the critical regime present in the decoupled limit. Notably, a detailed analysis of this scenario leads to a single-parameter tuned transition between two non-invertible SSB phases [Fig.~\ref{fig:main_results}(c)]. 
The field theory at the critical point is described by a Luttinger parameter $K <1$ with central charge $c = 2$ (due to the two copies), and the scaling dimensions of the order parameters in both phases are identical. 
There are three specific features of this critical point that are worth emphasizing.
First, in the vicinity of the critical point, there is an emergent continuous non-invertible ``cosine" symmetry, similar to the ones mentioned in Ref.\cite{chang2021lorentzian, thorngren2024fusion, eck2024from, cao2025global, seifnashri2025gauging}. 
Second, there is a further enhanced Kramers-Wannier-like self-duality at the critical point that exchanges the two order parameters up to a sign, thereby establishing that the order parameters in both phases must have the same scaling dimension. This is reminiscent of the self-duality of the DQCP between easy-plane Neel and VBS phases~\cite{motrunich2004emergent,senthil2004deconfined, senthil2004quantum,senthil2023deconfine}.
{
Third, since the order parameters of both the $(D^o D^e, S D^o D^e)$-breaking and $(S, D^o D^e)$-breaking phases transform trivially under $\eta^o$ and $\eta^e$, they remain local under any gauging scheme (including twisted variants) associated with these symmetries. This implies that the DQCPs we identified persist as DQCPs even after gauging, providing a pathway for generating a large family of related DQCPs.
}

The fact that the DQCP structure is preserved under the aforementioned set of gaugings allows us to map DQCP with non-invertible symmetries to DQCP with only invertible symmetries. The advantage of working with a system with only invertible symmetries is that the field-theoretic analysis (using bosonization) is significantly simplified, which allows us to fully map out the phase diagram in the vicinity of the critical point. The specific gauging procedure we employ is originally due to Baxter~\cite{baxter1972one, mahito1981hamiltonian}. Concretely, denoting $\eta^o = \prod_j X_{2j-1}$ and $\eta^e = \prod_j X_{2j}$ as the $\mathbb{Z}_2$ symmetry generators acting on the odd and even sites, respectively, we find that the Baxter's transformation maps the symmetry generators $(\eta^o, \eta^e, S D^o D^e, S)$ to $(\eta^X, \eta^Z, T, \eta^{\mathsf{H}})$. Here, $\eta^X = \prod_j X_j$, $\eta^Z = \prod_j Z_j$, and $\eta^{\mathsf{H}} = \prod_j \mathsf{H}_j$, where $X_j$ and $Z_j$ are Pauli matrices and $\mathsf{H}_j = (X_j + Z_j)/\sqrt{2}$. The operator $T$ denotes the lattice translation in the gauged model.
Phrased in more modern language, the Baxter transformation amounts to first gauging the $\mathbb{Z}_2^o$ symmetry, followed by gauging the diagonal subgroup $\tilde{\mathbb{Z}}_2^o \times \mathbb{Z}_2^e$, where $\tilde{\mathbb{Z}}_2^o$ is the dual of the $\mathbb{Z}_2^o$ symmetry. Mapping a model with a non-invertible symmetry $S D^o D^e$ to a model with translational symmetry is reminiscent of the fact that the Ising model with KW symmetry can be fermionized to Majorana fermions with translational symmetry. Interestingly, the $(D^o D^e, S D^o D^e)$-breaking phase we identify is  mapped to a translational symmetry-breaking phase—namely, the well-known valence bond solid (VBS) phase~\cite{majumdar1970antiferromagnetic,majumdar1969next}. This is analogous to the mapping between the KW SSB phase of the Ising model and the translational SSB phase of Majorana fermions~\cite{obrien2018lattice}.
{Furthermore, the gauged system is known to possess a Lieb-Schultz-Mattis (LSM)-type anomaly, offering a natural platform for realizing a DQCP. We remark that the mapping between a non-invertible symmetry and an anomalous invertible symmetry illustrates a general phenomena that gauging a non-anomalous subgroup in an anomalous system can result in a system with non-invertible symmetry.
 ~\cite{tachikawa2020gauging, thorngren2020anomalies, ji2020topological, kaidi2022kramers, seifnashri2024lieb,seifnashri2024cluster,choi2025non}.}

The Baxter transformation offers several advantages for understanding the models of interest in this paper. For example, the symmetry considered in Fig.\ref{fig:main_results}(a) is generated by $(\eta^o, \eta^e, S D^o D^e)$, which is expected to flow to the Rep($D_8$) fusion category at the decoupled Ising critical point \cite{thorngren2024fusion}. We note that another closely related symmetry described by Rep($D_8$) is generated by $(\eta^o, \eta^e, T^{-1} D^o D^e)$ and has been explored recently \cite{seifnashri2024cluster,li2024noninvertible}. However, unlike $T^{-1} D^o D^e$, the non-invertible symmetry $S D^o D^e$ mixes with the lattice translation and thus the symmetry considered here does not form a fusion category on the lattice. 
The Baxter transformation reinforces the expectation that the symmetry $(\eta^o, \eta^e, S D^o D^e)$ flows to its continuum counterpart at the decoupled Ising critical point. This is because the gauged model corresponds to the XY model, in which the $\mathbb{Z}_L$ translation emanates as a $\mathbb{Z}_2$ operator in the infrared. Consequently, in the original model, the relation $(S D^o D^e)^2 = T^o T^e (I + \eta^o)(I + \eta^e) \sim (I + \eta^o)(I + \eta^e)$ holds in the low-energy limit.
Furthermore, in the gauged model, the phase diagram near the XY fixed point with perturbation respecting $(\eta^X, \eta^Z, T)$  has been extensively studied both analytically and numerically~\cite{jiang2019ising, roberts2019deconfined, huang2019emergent}. Therefore, many established results can be directly applied to our system by keeping track of the global properties of the Baxter transformation.

To underline unexplored aspects, we will primarily focus on the scenario in Fig.\ref{fig:main_results}(b)[i.e., imposing the symmetry generators $(\eta^o, \eta^e, S D^o D^e, S)$] and Fig.\ref{fig:main_results}(c), while briefly discussing the scenario in Fig.\ref{fig:main_results}(a) along the way.
The Baxter transformation will continue to be helpful in identifying distinct phases and their transitions.
{
We note that the symmetry generators $(\eta^o, \eta^e, S D^o D^e, S)$ enforced in Fig.~\ref{fig:main_results}(b) form the union of two subsets: $(\eta^o, \eta^e, S D^o D^e)$ and $(\eta^o, \eta^e, D^o D^e)$. Near the decoupled Ising critical point, the former flows to $\text{Rep}(D_8)$, while the latter flows to $\text{Rep}(H_8)$~\cite{thorngren2024fusion}.
}
Both fusion categories are anomaly-free: $\text{Rep}(D_8)$ admits three distinct symmetric phases, while $\text{Rep}(H_8)$ admits only one~\cite{thorngren2024fusion}. It is therefore natural to expect that the system we consider is also anomaly-free. We will verify this expectation in Sec.~\ref{sec:second_scenario} by explicitly constructing a lattice model that respects the symmetry $(\eta^o, \eta^e, S D^o D^e, S)$ and possesses a unique gapped ground state.
However, this does not contradict with the aforementioned single-parameter tuned transition between two non-invertible SSB phases, as anomalies can also emerge in the vicinity of critical points or within certain phases \cite{metlitski2018intrinsic}. In fact, we can map our DQCP via gauging to another system that also shows DQCP (for invertible symmetries) and which has \textit{intrinsic} LSM anomaly (Sec.\ref{sec:gauging}). 
This leads us to conjecture that an intrinsic anomaly in the gauged system can manifest as an emergent anomaly in the original (i.e., ungauged) system. For our problem of interest, such an emergent anomaly would then forbid the existence of a symmetric gapped phase in the vicinity of the DQCP critical point.

The rest of the paper is organized as follows. Sec.~\ref{sec:second_scenario} analyzes a $\mathbb{Z}_2^o \times \mathbb{Z}_2^e$ symmetric spin chain with additional $S D^o D^e$ and $S$ symmetries. We identify the resulting gapped phases, characterize their order parameters, and study the associated phase transitions using the Baxter transformation. We also briefly consider the case where only $S D^o D^e$ is imposed as an additional symmetry. Sec.~\ref{sec:two_copy} introduces two coupled chains that realize a continuous transition between non-invertible symmetry-breaking phases. We develop the corresponding critical theory in an alternative set of field variables, uncover an emergent cosine symmetry and self-duality, and demonstrate how gauging spin-flip symmetries generates new families of DQCPs. Sec.~\ref{sec:outlook} concludes with a summary and outlook. App.~\ref{sec:Baxter} computes the ground-state degeneracies of the gapped phases in Sec.~\ref{sec:second_scenario} via the Baxter transformation.

\section{$\mathbb{Z}^o_2 \times \mathbb{Z}^e_2$ spin chain with $S D^o D^e$ and $S$ symmetry}
\label{sec:second_scenario}

The Hilbert space of our system consists of spin-1/2 degrees of freedom on each vertex of a 1d lattice. We further distinguish odd and even sites, and for any operator $A$, label $A_{2j-1}$ as $A^o_{j}$ and $A_{2j}$ as $A^e_{j}$.
We consider the class of Hamiltonians with symmetry generators $(\eta^o, \eta^e, S,  S D^o D^e)$. Here $\eta^o = \prod_j X_{2j-1}$, $\eta^e = \prod_j X_{2j}$, $D^o$ and $D^e$ generate the KW transformation on odd and even sites respectively, and $S = \prod_j (I +X_{2j-1} X_{2j} +Y_{2j-1} Y_{2j} + Z_{2j-1} Z_{2j})/2$ is the swap operator that implements $A_{2j-1} \leftrightarrow A_{2j}$ for any operator $A$. We now investigate phases and phase transitions of this class of Hamiltonians close to the decoupled Ising critical point:
\begin{equation}
\label{Eq:second_scenario}
\begin{aligned}
    {H} =  & -\sum_{j=1}^{L}  \Big[X^{o}_j+Z^{o}_j Z^{o}_{j+1} + (o \leftrightarrow e)\Big] +\delta {H}_{\text{int}},
\end{aligned}
\end{equation}
where $\delta {H}_{\text{int}}$ is any perturbation symmetric under $\eta^o, \eta^e$, $S D^o D^e$, and $S$. 
{
An example of $\delta H_{\text{int}}$ is $\delta H_{\text{int}} = \lambda \sum_j Z^o_j Z^o_{j+1}(X^e_j + X^e_{j+1}) + (o \leftrightarrow e)$, which, as discussed in Sec.\ref{sec:identify_phases}, drives the system into a non-invertible SSB phase with a frustration-free fixed point at $\lambda = 1/2$ (see Eq.~\eqref{Eq:SD_fixedpoint_ham}). Another example is $\delta H_{\text{int}} = \lambda' \sum_j Z^o_j Z^o_{j+1} Z^e_{j+1} Z^e_{j+2} + X^o_j X^e_{j+1} + (o \leftrightarrow e)$ which drives the system into a different non-invertible SSB phase with a frustration-free fixed point at $\lambda' = 1/2$ (see Eq.~\eqref{Eq:SandD_fixedpoint_ham}).
}

Our main results in this section are as follows. We identify four gapped phases near the fixed point (see Table \ref{table:2}): the partially ordered phase (which breaks both $\eta^o$ and $\eta^e$ but preserves $\eta^o \eta^e$), the $(D^o D^e, S D^o D^e)$-breaking phase (which breaks both $S D^o D^e$ and $D^o D^e$ while preserving $S$), the symmetric phase, and the $(S, D^o D^e)$-breaking phase (which breaks both $D^o D^e$ and $S$ while preserving $S D^o D^e$). {We will present the fixed-point Hamiltonians of all phases in Sec.~\ref{sec:identify_phases}.
}
The partially ordered phase can be understood using conventional symmetry-breaking mechanisms, with a local order parameter $Z^o_j Z^e_j$.
On the other hand, the $(D^o D^e, S D^o D^e)$-breaking and $(S, D^o D^e)$-breaking phases cannot be explained using ordinary symmetries and can only be understood as the SSB of the non-invertible symmetries. 
The corresponding local order parameter for these non-invertible SSB phases are  $\mathcal{O}_j = (X^o_j - Z^o_j Z^o_{j+1}) + (o \leftrightarrow e)$ and $\mathcal{Q}_j = (X^o_j - Z^o_j Z^o_{j+1}) -(o \leftrightarrow e)$, respectively.
Single-parameter tuned transitions exist between the partially ordered phase and the $(D^o D^e, S D^o D^e)$-breaking phase, as well as between the symmetric phase and the $(S, D^o D^e)$-breaking phase. However, these transitions are not regarded as DQCPs, as the partially ordered phase and the symmetric phase each respect more symmetries than the $(D^o D^e, S D^o D^e)$-breaking and $(S, D^o D^e)$-breaking phases, respectively. On the other hand, the two non-invertible SSB phases are separated by a critical regime [see Fig.\ref{fig:main_results}(b)]. Therefore, no DQCP exists in the current setup. However, we will show in Sec.~\ref{sec:two_copy} that a direct transition between analogous phases can be achieved by introducing an additional copy of a $\mathbb{Z}_2 \times \mathbb{Z}_2$ symmetric spin chain with symmetry-allowed inter-chain interactions.
{
We also briefly examine the scenario in Fig.~\ref{fig:main_results}(a), in which the symmetry generators include $(\eta^o, \eta^e, S D^o D^e, S)$, and show that there is a single-parameter tuned transition between two conventional (group-like) SSB phases.
}

\begin{figure}
\centering
\includegraphics[width=\linewidth]{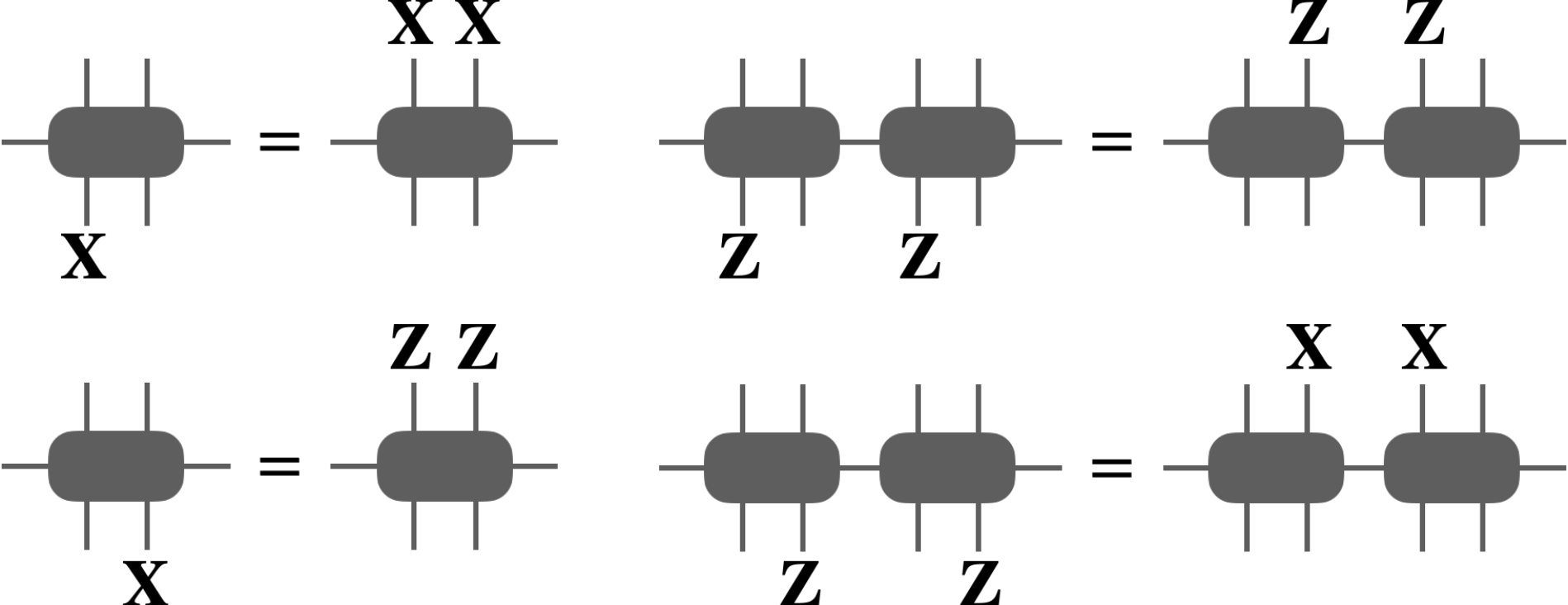}
\caption{Schematic representation of the Baxter transformation in Eq.\eqref{Eq:baxter_trans} (time flows upward).
}
\label{fig:baxter_trans}
\end{figure}

\subsection{Mapping to a system with invertible symmetry via Baxter transformation}
\label{sec:baxter_short}

To identify the gapped phases and their phase transitions close to the Ising critical fixed point, it is expedient to use the Baxter transformation \cite{baxter1972one, mahito1981hamiltonian} that gauges our system to one that possesses only invertible (i.e. group-like) symmetries. We will show below that the symmetry generators in the gauged system are $\eta^X = \prod_j X_j, \eta^Z = \prod_j Z_j $, $\eta^{\mathsf{H}} = \prod_j \mathsf{H}_j $, and $T$, where $T$ is the translational operator and $\mathsf{H}_j = (X_j + Z_j)/\sqrt{2}$ is the Hadamard gate.
The operator $B$ that implements the Baxter transformation takes the form
\begin{equation}
\label{Eq:baxter_explicit}
B = D^\dagger D^o,
\end{equation}
where $ D^o$ is the KW transformation on the odd sites (i.e., $ D^o$ sends $X_{2j-1}$ to $Z_{2j-1}Z_{2j+1}$,  $Z_{2j-1}Z_{2j+1}$ to $X_{2j+1}$, while leaving all the even-site operators unchanged) and $D$ is the KW transformation on all sites (i.e., $ D$ sends $X_{j}$ to $Z_{j}Z_{j+1}$ and  $Z_{j}Z_{j+1}$ to $X_{j+1}$). 
Recall that any $\mathbb{Z}^o_2 \times \mathbb{Z}^{e}_2$ symmetric Hamiltonian is generated by the local operators $X^o_j, X^e_j, Z^o_j Z^o_{j+1}$, and $Z^e_j Z^e_{j+1}$. 
A straightforward use of Eq.\eqref{Eq:baxter_explicit} implies
$B$ implements the following transformation on these generators:
\begin{equation}
\label{Eq:baxter_trans}
    \begin{aligned}
        X^o_{j} &= X_{2j-1}  & \rightsquigarrow  X_{2j-1} X_{2j} ,  \\
        X^e_{j} &= X_{2j} & \rightsquigarrow  Z_{2j-1} Z_{2j} ,  \\
         Z^o_{j}Z^o_{ j+1} &= Z_{2j-1} Z_{2j+1} &\rightsquigarrow  Z_{2j} Z_{2j+1} ,  \\
         Z^e_{j}Z^e_{j+1} & = Z_{2j} Z_{2j+2}  & \rightsquigarrow  X_{2j} X_{2j+1},  \\
    \end{aligned}
\end{equation}
where  $O \rightsquigarrow Q$ is a shorthand notation for $B O = Q B$ (see Fig.\ref{fig:baxter_trans} for a schematic representation). Note that we intentionally separate the original model (i.e., the model before applying the Baxter transformation) into even and odd sites, whereas we do not apply this separation to the gauged model. This distinction helps clarify whether we are referring to the original or gauged model later.

To understand the topological properties of the mapping, we note that the first two lines in Eq.\eqref{Eq:baxter_trans} imply
\begin{equation}
\label{Eq:uoe_uxz}
\begin{aligned}
 \eta^o = \prod_j X^o_j & \rightsquigarrow \eta^X = \prod_j X_j, \\
 \eta^e = \prod_j X^e_j & \rightsquigarrow \eta^Z = \prod_j Z_j,
\end{aligned}    
\end{equation}
and hence $B$ maps a $\mathbb{Z}_2^o \times \mathbb{Z}_2^e$ symmetric spin chain to a $\mathbb{Z}_2^X \times \mathbb{Z}_2^Z$ symmetric spin chain.
On the other hand, the last two lines in Eq.\eqref{Eq:baxter_trans} imply
\begin{equation}
\label{Eq:toe_uxz}
\begin{aligned}
(-1)^{t^o} & \rightsquigarrow (-1)^{t^X} \eta^Z, \\
(-1)^{t^e} & \rightsquigarrow (-1)^{t^Z} \eta^X,
\end{aligned}
\end{equation}
Eqs.\eqref{Eq:uoe_uxz}  and \eqref{Eq:toe_uxz} imply that the $\mathbb{Z}_2$-charged sector $(u^o, u^e)$ and the $\mathbb{Z}_2$-twisted 
sector $(t^o, t^e)$ of the original theory are mapped to the $\mathbb{Z}_2$-charged sector 
$(u^X, u^Z)$ and the $\mathbb{Z}_2$-twisted sector $(t^X, t^Z)$ of the dual theory in a 
nontrivial way, where $\eta^{o/e} = (-1)^{u^{o/e}}$ and $\eta^{X/Z} = (-1)^{u^{X/Z}}$.
Specifically, $B$ maps the system in the $\mathbb{Z}_2$-charged sector $(u^o, u^e)$ and $\mathbb{Z}_2$-twisted sector $(t^o, t^e)$  to the $\mathbb{Z}_2$-charged sector $(u^X,u^Z ) = (u^o, u^e)$ and the $\mathbb{Z}_2$-twisted sector $(t^X, t^Z) = (t^o+u^e, t^e+u^o)$, respectively. 
Note that while $B$ is non-invertible over the entire Hilbert space, it acts as a unitary 
operator in each given $\mathbb{Z}_2$-charged sector $(u^o, u^e)$ and $\mathbb{Z}_2$-twisted 
sector $(t^o, t^e)$.
These properties are essential for deducing the total ground-state degeneracy (GSD) $d^{(t^o, t^e)}_{\text{total}}$ in any twisted sector $(t^o, t^e)$ of the original model, based on those of the gauged system, as elaborated in App.~\ref{sec:Baxter}.

It is also insightful to examine  the action  of the Baxter transformation on the symmetry generators $S D^o D^e$ and $S$. Eq.\eqref{Eq:baxter_trans} implies
\begin{equation}
\begin{aligned}
 S D^o D^e & \rightsquigarrow T, \\
 S  & \rightsquigarrow \eta^\mathsf{H}, \\
\end{aligned}
\end{equation}
Therefore, the gauged model  respects the symmetry generators $(\eta^X, \eta^Z, T, \eta^{\mathsf{H}} )$.
Since the unit cell of the gauged model transforms projectively under the onsite symmetries, it satisfies the LSM constraint that rules out a symmetric gapped phase. However, this does not necessarily imply that the original model possesses an intrinsic LSM anomaly, as the Baxter transformation maps some local (non-local) operators to non-local (local) operators.
{

Our overall strategy for studying the model in Eq.~\eqref{Eq:second_scenario} is to first analyze the phase diagram and critical behavior of the gauged models near the transition points using effective field theory. We then reverse the Baxter transformation to investigate the corresponding features in the original model. Note that since $B$ is non-invertible over the full Hilbert space, by reversing the Baxter transformation, we refer specifically to restricting the Hilbert space to a fixed $\mathbb{Z}_2$-charged and $\mathbb{Z}_2$-twisted sector, within which $B$ becomes invertible.

Using Eq.~\eqref{Eq:baxter_trans}, the decoupled Ising critical point is mapped to the XY model, 
which is well known \cite{mahito1981hamiltonian}.
Therefore, applying the Baxter transformation yields the following Hamiltonian
\begin{equation}
\label{Eq:second_scenario_dual}
\begin{aligned}
    \tilde{H} =  & -\sum_{j=1}^{2L}  (X_j X_{j+1}+ Z_j Z_{j+1}) +\delta \tilde{H}_{\text{int}},
\end{aligned}
\end{equation}
where $\delta \tilde{H}_{\text{int}}$ denotes all possible perturbations symmetric under $\eta^X, \eta^Z$, $T$, and $\eta^{\mathsf{H}}$. This is analogous to the model exhibiting DQCP studied in Ref.~\cite{jiang2019ising} with additional symmetry $\eta^{\mathsf{H}}$.
Therefore, we will use the same bosonization technique as in Ref.~\cite{jiang2019ising} to study our system. 
Specifically, the Pauli operators act in the continuum limit as follows:
\begin{equation}
\label{Eq:lattice_continuum}
    \begin{aligned}
        Z_j & \sim \cos(\phi) \\
        X_j & \sim -\sin(\phi) \\
        Y_j & \sim \frac{\partial_x \theta}{\pi} + A(-1)^j \sin(\theta),
    \end{aligned}
\end{equation}
where $\theta$ is the dual of $\phi$ and they satisfy $[\partial_x \theta(x), \phi(x)] = i 2\pi \delta(x-x')$. 
We note that Eq.\eqref{Eq:lattice_continuum} implies both $\phi$ and $\theta$ are $2 \pi$ periodic.
The transformation properties of the fields $(\phi, \theta)$ under the symmetry generators $(\eta^X, \eta^Z, T, \eta^{\mathsf{H}})$ can be determined by first recalling how these symmetry generators 
act on the Pauli operators and then applying Eq.~\eqref{Eq:lattice_continuum}. The results are:
\begin{equation}
\label{Eq:sym_continuum}
    \begin{aligned}
    T:\ & (\phi, \theta ) \rightarrow (\phi, \theta + {\pi}), \\
    \eta^Z:\ & (\phi, \theta ) \rightarrow (-\phi, -\theta ), \\ 
    \eta^X:\ & (\phi, \theta ) \rightarrow (-\phi + \pi, -\theta), \\    
        \eta^{\mathsf{H}}:\ & (\phi, \theta ) \rightarrow (-\phi - \frac{\pi}{2}, -\theta).
    \end{aligned}
\end{equation}
Note that the translation, which corresponds to a $\mathbb{Z}_L$ symmetry on the lattice, effectively behaves as a $\mathbb{Z}_2$ symmetry in the low-energy limit. This is because all low-energy states $|\tilde{\psi}\rangle$ near the Gaussian fixed point satisfy $T^2|\tilde{\psi}\rangle = |\tilde{\psi}\rangle$. This property is well-known and has recently been highlighted in Ref.~\cite{metlitski2018intrinsic,cheng2023lieb} in the context of emergent anomalies. Following the convention in Ref.~\cite{cheng2023lieb}, we refer to this as the 
\textit{emanant} symmetry to distinguish it from the emergent symmetry, which refers to a 
symmetry that becomes enlarged in the low-energy limit, as all interactions violating the 
enlarged symmetry are irrelevant in the renormalization group sense.
As we will see in the following sections, this emanant symmetry plays a crucial role in identifying the order parameters of the non-invertible SSB phases.
Now, from Eq.\eqref{Eq:sym_continuum}, the symmetry-preserving action can be written as 
\begin{equation}
\label{Eq:field_SDoDe}
\begin{aligned}
 S[\phi &, \theta] =  S_0  + S_{\text{int}},
\end{aligned}
\end{equation}
where 
\begin{equation}
\label{Eq:S_0}
S_0 = \int d\tau d x   \Big[ \frac{i}{2 \pi} \partial_\tau \phi  \partial_x \theta + \frac{v}{2\pi} \Big(\frac{1}{4 g}(\partial_x \theta)^2 + g (\partial_x \phi)^2 \Big) \Big],
\end{equation}
and
\begin{equation}
S_{\text{int} }= \int d\tau d x  \big[\lambda_\theta \cos(2\theta) +\lambda_\phi \cos(4 \phi) + \cdots \big],
\end{equation}
and $\cdots$ denotes the less relevant terms. 

\subsection{Identification of phases}
\label{sec:identify_phases}

\begin{table*}
\renewcommand{\arraystretch}{1.5}
\begin{tabular}{|c | c| c| c| c|} 
 \hline  \hline
Original Model   & $\lambda_\theta \rightarrow \infty $& $\lambda_\theta \rightarrow -\infty$ & $\lambda_\phi \rightarrow \infty$ & $\lambda_\phi \rightarrow - \infty$ \\ [0.5 ex] 
 \hline \hline
 Phase  & partially ordered & $(D^o D^e, S D^o D^e)$-breaking & Symmetric & $(S, D^o D^e)$-breaking\\
 \hline
  fixed-point wave functions & \makecell{$\prod_j  |\phi^+_{(o,j), (e, j)} \rangle$,  \\
 $\prod_j |\psi^+_{(o,j), (e, j)} \rangle$} & \makecell{$|x,+\rangle_o |x,+\rangle_e$, \\ $|z,\pm \rangle_o |z,\pm\rangle_e$}  & $B^\dagger (I+\eta^X)(I+\eta^Z)|h,+\rangle$ & \makecell{$|x,+\rangle_o |z,\pm\rangle_e$, \\ $|z,\pm \rangle_o |x,+\rangle_e$} \\
\hline
\makecell{GSD  under PBC in $ (u^o, u^e)$\\ $ = (0,0), (0,1), (1,0), (1,1)$}  & $2 = 1+0+0+1$ & $5 = 2+1+1+1$ & $1 = 1+0 + 0 + 0$ & $4 = 2+1+1+0$ \\
\hline
(dis)order parameter & \makecell{$i Z^o_j Z^e_j   (X^o_j X^e_j - I) \times $ \\ $ (\prod_{k\geq j+1} X^o_j X^e_j) $} & \makecell{$(X^o_j -Z^o_j Z^o_{j+1} ) + $ \\$( X^e_j- Z^e_j Z^e_{j+1}) $} & 
\makecell{$Z^o_j Z^e_j (X^o_j + X^e_j) \times$ \\ $  (\prod_{k\geq j+1} X^o_k X^e_k)$  }
& \makecell{$(X^o_j - Z^o_j Z^o_{j+1}) - $ \\$( X^e_j - Z^e_j Z^e_{j+1} )$}  \\
\hline \hline
  Gauged Model & $\lambda_\theta \rightarrow \infty $& $\lambda_\theta \rightarrow -\infty$  & $\lambda_\phi \rightarrow \infty$ & $\lambda_\phi \rightarrow - \infty$  \\ [0.5 ex] 
\hline \hline
Phase & $y$-AFM & VBS & $(\eta^Z, \eta^X)$-breaking & $(\eta^{\mathsf{H}, }, \eta^{\mathsf{H^-}})$-breaking  \\
 \hline
  fixed-point wave functions   &  $\prod_j(|y,\pm \rangle_{2j-1} |y,\mp \rangle_{2j}) $  & \makecell{$\prod_j |\phi^+_{2j-1, 2j}\rangle $,\\ $T(\prod_j |\phi^+_{2j-1, 2j}\rangle)$} & \makecell{$|h,\pm \rangle$,  \\ $\eta^X|h,\pm \rangle$} & \makecell{$|z,\pm \rangle$, \\ $|x,\pm \rangle$} \\
\hline
\makecell{GSD  under PBC in $ (u^X, u^Z)$\\ $ = (0,0), (0,1), (1,0), (1,1)$} & $2 = 1+0+0+1$ & $2=2+0+0+0$ & $4 = 1+1+1+1$ & $4 = 2+1+1+0$  \\
\hline
order parameter & $\sin(\theta) \sim Y_j - Y_{j+1}$ &\makecell{ $\cos(\theta) \sim$ \\$(Z_{j-1} Z_{j}+X_{j-1} X_{j}) -$\\ $ (Z_{j} Z_{j+1}+X_{j} X_{j+1}) $ } & \makecell{ $\sin(2\phi) \sim$ \\$Z_j X_{j+1} + X_j Z_{j+1}$} & \makecell{$\cos(2\phi) \sim$ \\ $ X_j X_{j+1} - Z_j Z_{j+1} $}  \\
\hline
\end{tabular}
\caption{Properties of the gapped phases described by the original Hamiltonian in Eq.\eqref{Eq:second_scenario} and its gauged Hamiltonian in  Eq.\eqref{Eq:second_scenario_dual}. The field theory of the gauged model is described by a compact free bosons with interactions
$S_{\text{int}  } = \int d\tau d x  \big[\lambda_\theta \cos(2\theta) + \lambda_\phi \cos(4 \phi)  \big].$
{
Here, $|z,\pm\rangle = \prod_j |z,\pm\rangle_j$ denotes a product state in the Pauli-$Z$ basis with positive or negative eigenvalue, and similar notation applies to $|x,\pm\rangle$. Similarly, $|h,\pm\rangle$ denotes a product state in the Hadamard basis, where $\mathsf{H} = (X + Z)/\sqrt{2}$. Finally, $|\phi^+_{j,j+1}\rangle \propto |0_j 0_{j+1}\rangle + |1_j 1_{j+1}\rangle$, $|\psi^+_{j,j+1}\rangle \propto |0_j 1_{j+1}\rangle + |1_j 0_{j+1}\rangle$, and $B = D^\dagger D^o$ denotes the Baxter transformation (see Eq.~\eqref{Eq:baxter_explicit}).
}
}
\label{table:2}
\end{table*}

We will now identify gapped phases described by this field theory and then reverse the Baxter transformation to study the original model. The results are summarized in Table.\ref{table:2}.
Our strategy is to first write down the fixed-point Hamiltonian and wavefunctions of each phase.
We will then elaborate on the properties of the original model by reversing the Baxter transformation. In particular, we will write down the fixed-point Hamiltonian, determine the ground-state degeneracies, and identify the order (or disorder) parameters of all the gapped phases of the original model.

\begin{itemize}
\item $\lambda_\theta \rightarrow \infty$:  \textbf{partially ordered phase}
\\
To minimize the action, $\theta$ is pinned to $\pi/2$ or $3\pi/2$, and the corresponding fixed-point wave functions are $\prod_j(|y,+\rangle_{2j-1} |y,- \rangle_{2j}) $ and $\prod_j (|y,-\rangle_{2j-1} |y,+ \rangle_{2j})$, respectively. 
{Here $|y, \pm\rangle $ denotes the eigenvector of the Pauli-$Y$ matrix with positive or negative eigenvalues.}
We identify this as the $y$-AFM phase.
The order parameter that takes non-zero value in all physical ground states is $\sin( \theta)$, which corresponds to $Y_j - Y_{j+1}$ in the UV.
The fixed-point Hamiltonian is $\tilde{H} = \sum_j Y_j Y_{j+1}$.
The original fixed-point Hamiltonian obtained by undoing the Baxter transformation is $ H = -\sum_j (X^o_{j} X^e_{j}  + Z^o_{j} Z^o_{j+1} Z^e_{j} Z^e_{j+1} ).$
Therefore, the ground-state subspace is two-dimensional and is spanned by $\prod_j  |\phi^+_{(o,j), (e, j)} \rangle$ and $\prod_j |\psi^+_{(o,j), (e, j)} \rangle$, where $|\phi^+\rangle \propto |00\rangle + |11\rangle$, $|\psi^+\rangle \propto |01\rangle + |10\rangle$. It preserves $\eta^o \eta^e$  but breaks $\eta^o$ and $\eta^e$. We denote this as a partially order phase following Ref.\cite{mahito1981hamiltonian}. The local order parameter for the $\eta^o$-breaking phase is identified as $Z^o_j Z^e_j$, 
which carries charge under $\eta^o$ and $\eta^e$ but remains neutral under $\eta^o \eta^e$, 
corresponding to $\prod_{k \geq 2j} Y_k$ in the gauged model.

\item \textbf{$\lambda_\theta \rightarrow -\infty $: $(D^o D^e, S D^o D^e)$-breaking phase}.
\\
To minimize the action, $\theta$ is pinned to $0$ or $\pi$, and the corresponding fixed-point wave functions can be identified as $\prod_j |\phi^+_{2j-1, 2j}\rangle $ and  $\prod_j |\phi^+_{2j, 2j+1}\rangle $, respectively.
This corresponds to the well-known VBS phase. The order parameter that acquires a nonzero value in this phase is $\cos(\theta)$, which emanates from the following local operator in the UV:
\begin{equation}
\label{Eq:order_VBS}
\tilde{\mathcal{O}}_j = (Z_{j-1} Z_{j} + X_{j-1} X_{j}) - ( j \rightarrow j+1).
\end{equation}
Note that under the translation, $\tilde{\mathcal{O}}_j$ transforms as $\tilde{\mathcal{O}}_j \rightarrow \tilde{\mathcal{O}}'_j = -[(Z_{j+1} Z_{j+2} + X_{j+1} X_{j+2}) - (Z_{j} Z_{j+1} + X_{j} X_{j+1})] \neq -\tilde{\mathcal{O}}_j$, indicating that it does not transform in a simple manner under the $\mathbb{Z}_L$ translational symmetry. However, due to the aforementioned emanant $\mathbb{Z}_2$ symmetry, all low-energy states $|\tilde{\psi}\rangle$ satisfy $T^2 |\tilde{\psi}\rangle = |\tilde{\psi}\rangle$, implying that the effect of $\tilde{\mathcal{O}}'_j$ on any low-energy state is exactly the same as that of $-\tilde{\mathcal{O}}_j$ on any low-energy state. The fixed-point Hamiltonian of the VBS phase is 
\begin{equation}
\label{Eq:VBS_ham}
\begin{aligned}
\tilde{H} & =\sum_j (I - Z_j Z_{j+1})(I - Z_{j+1} Z_{j+2}) + (Z \leftrightarrow X).
\end{aligned}
\end{equation}
Unlike the partially ordered phase, while Eq.~\eqref{Eq:VBS_ham} is written as a sum of projectors, 
the local terms do not commute with one another. However, Refs.~\cite{majumdar1970antiferromagnetic, 
majumdar1969next} have shown that the Hamiltonian is gapped and possesses the aforementioned 
two-fold degenerate ground states.
The original Hamiltonian obtained by undoing the Baxter transformation is
\begin{equation}
\label{Eq:SD_fixedpoint_ham}
\begin{aligned}
 {H} =  \sum_j (I - Z^o_{j} Z^o_{j+1})[(I - X^e_{j} ) + (I & - X^e_{j+1})]
  \\
 & +(o \leftrightarrow e) .
\end{aligned}
\end{equation}
One can easily see that the ground-state subspace includes $|x,+\rangle_o |x,+\rangle_e$, $|z,\pm \rangle_o |z,\pm\rangle_e$, as they are all eigenvectors of $H$ with eigenvalues $0$.
{Here, $|z,\pm\rangle$ denotes a product state in the Pauli-$Z$ basis with positive or negative eigenvalue, and similar notation applies to $|x,\pm\rangle$ in the Pauli-$X$ basis.
}
The fact that these five states are the only ground states can be shown in various ways, including similar arguments in Ref.\cite{obrien2018lattice, gorantla2024duality, zhang2024long}.
In App.~\ref{sec:Baxter}, we use the topological property of the Baxter transformation—namely, that $B$ maps a system in the $\mathbb{Z}_2$-charged sector $(u^o, u^e)$ and the $\mathbb{Z}_2$-twisted sector $(t^o, t^e)$ to the charged sector $(u^X, u^Z) = (u^o, u^e)$ and the twisted sector $(t^X, t^Z) = (t^o + u^e, t^e + u^o)$—to demonstrate the five-fold ground-state degeneracy.
{We refer to this phase as the $(D^o D^e, S D^o D^e)$-breaking phase, as it breaks both the $S D^o D^e$ and $D^o D^e$ symmetries. This can be seen by noting that there does not exist a physical ground state that is invariant under these symmetries—that is, all physical states have vanishing expectation values with respect to the corresponding symmetry operators. Here, we follow the terminology in Ref.\cite{bhardwaj2025gapped, bhardwaj2024illustrating, bhardwaj2024lattice, chatterjee2024quantum} to call a state $|\psi\rangle$ spontaneously breaks a non-invertible symmetry $W$ if its expectation value $\langle \psi |W |\psi\rangle$ is vanishing.
}
We note that another exactly solvable fixed-point Hamiltonian, whose five-fold degenerate ground states are $|x,+\rangle_o |x,+\rangle_e$ and $|z,\pm \rangle_o |z,\pm\rangle_e$, 
was constructed in Ref.~\cite{zhang2024long}. However, the model in Ref.~\cite{zhang2024long} differs from the one constructed here: the model we considered possesses swap symmetry but not single-site translational symmetry (characterized by $O^o_j \rightarrow O^{e}_j,\ O^e_j \rightarrow O^o_{j+1}$ for any operator $O$), whereas the model in Ref.~\cite{zhang2024long} possesses translational symmetry but breaks swap symmetry.

We now identify the local order parameter that characterizes the $(D^o D^e, S D^o D^e)$-breaking phase.  Interestingly, since $\tilde{\mathcal{O}}_j$ in Eq.\eqref{Eq:order_VBS} transforms trivially under both $\eta^X$ and $\eta^Z$, it is also the local order parameter after reversing the Baxter transformation. Specifically, the original order parameter takes the form
\begin{equation}
\mathcal{O}^{}_{j} = (X^o_j - Z^o_j Z^o_{j+1})  + ( o \leftrightarrow e).
\end{equation}
Similar to how $\tilde{\mathcal{O}}_j$ transforms under $T$ in the gauged model, $\mathcal{O}^{}_{j}$ does not transform straightforwardly under the $S D^o D^e$ symmetry, i.e., $\mathcal{O}_{j} \rightarrow (Z^o_j Z^o_{j+1} - X^o_{j+1}) + ( o \leftrightarrow e) \neq -\mathcal{O}_j$. However, in the low-energy limit, $\mathcal{O}^{}_{j}$ carries a charge under $S D^o D^e$ in the $(u_o, u_e) = (0,0)$ sector (note that $S D^o D^e |\psi\rangle =0$ if $|\psi\rangle$ is not in the $(u_o, u_e) = (0,0)$ sector). 
This follows from the fact that any low-energy state $|\psi\rangle$ in the $(u_o, u_e) = (0,0)$ sector satisfies $(S D^o D^e)^2|\psi\rangle = (I+\eta^o)(I+\eta^e)T^oT^e|\psi\rangle =|\psi\rangle$. This implies the effect of $X^o_{j+1} + X^e_{j+1}$ on any low-energy state is equivalent to the effect of $X^o_{j} + X^e_{j}$, and thus $\mathcal{O}_j$ is effectively charged under $S D^o D^e$.
One can similarly show that $\mathcal{O}_j$ is charged under $D^o D^e$ but remains invariant under $S$. These properties establish $\mathcal{O}_{j}$ is a local order parameter that characterizes the $(D^o D^e, S D^o D^e)$-breaking phase.

\item $\lambda_\phi \rightarrow \infty$: \textbf{symmetric phase}.
\\
$\phi$ is pinned to $\pi/4, 3\pi/4, 5\pi/4$ or $7\pi/4$, and the corresponding fixed-point wave functions are identified as [using Eq.\eqref{Eq:lattice_continuum}]  \( \eta_Z \ket{\mathsf{h},+} \), \( \eta_Z \eta_X \ket{\mathsf{h},+} \), \( \eta_X \ket{\mathsf{h},+} \), and \( \ket{\mathsf{h},+} \), respectively.
{Since all physical ground states transform non-trivially under $\eta^Z$ and $\eta^X$, the phase completely spontaneously breaks the $\mathbb{Z}_2 \times \mathbb{Z}_2$ subgroup generated by $(\eta^Z, \eta^X)$, and we refer to it as the $(\eta^Z, \eta^X)$-breaking phase.}
The order parameter that takes non-zero value for all physical ground states is $\sin(2\phi)$, which corresponds to $ Z_j X_{j+1} + X_j Z_{j+1} $ in the UV. The fixed-point Hamiltonian takes the form 
\begin{equation}
\tilde{H} = \sum_j (I- \mathsf{H}_j \mathsf{H}_{j+1})( I - \mathsf{H}^-_{j+2} \mathsf{H}^-_{j+3}) + (\mathsf{H} \leftrightarrow \mathsf{H^-}),
\end{equation}
where $\mathsf{H}^-_j = \eta_Z \mathsf{H}_j \eta_Z = (Z_j-X_j)/\sqrt{2}$.
Similar to the VBS phase, the fixed-point Hamiltonian cannot be written as a sum of commuting projectors. One can similarly verify that $\tilde{H}$ has a finite gap with aforementioned wave functions as its only ground states.
While the original Hamiltonian can be obtained by reversing the Baxter transformation, its explicit form is not simple and not particularly illuminating. However, the ground state of this phase can be shown to be unique using the topological properties of the Baxter transformation, as elaborated in App.~\ref{sec:Baxter}. 
{This explicitly shows that the system with symmetry generators $(\eta^o ,\eta^e, S, S D^o D^e)$ has no intrinsic anomaly.}
Furthermore, we show in App.~\ref{sec:Baxter} that the ground state is $\mathbb{Z}^e_2$-charged under a $\mathbb{Z}^o_2$ twist and vice versa, which is a hallmark of a $\mathbb{Z}^o_2 \times \mathbb{Z}^e_2$ SPT phase. We remark that although the symmetric phase resembles the cluster state with respect to the $\eta^o$ and $\eta^e$ symmetries, it is not adiabatically connected to the cluster state in the presence of the swap symmetry, since the cluster state does not respect the swap symmetry.

\item $\lambda_\phi \rightarrow -\infty$: \textbf{$(S, D^o D^e)$-breaking phase}.
\\
$\phi$ is pinned to $0, \pi/2, \pi$, or $3\pi/2$, and the corresponding fixed-point wave functions can be identified as $\ket{z,+}, \ket{x,-}, \ket{z,-}$, and $\ket{x,+}$, respectively.
It is apparent that this phase is a coexistence  between the $\eta^X$-breaking and $\eta^Z$-breaking phases, which is generally a first-order line but has become a stable phase due to the enforcement of $\eta^{\mathsf{H}}$ symmetry.
Since all the physical ground states transform non-trivially under $\eta^{\mathsf{H}}$ and $\eta^{\mathsf{H^-}} = \prod_j \mathsf{H}^-_j$, we call this phase the $(\eta^{\mathsf{H}}, \eta^{\mathsf{H^-}})$-breaking phase.
{We note that, similar to the $(\eta^Z, \eta^X)$-breaking phase, this phase spontaneously breaks the $\mathbb{Z}_2 \times \mathbb{Z}_2$ subgroup, though the subgroup is now generated by $(\eta^{\mathsf{H}}, \eta^{\mathsf{H}^-})$ instead of $(\eta^Z, \eta^X)$. This implies that this phase is distinct from the $(\eta^Z, \eta^X)$-breaking phase.
}
The order parameter that takes non-zero value for all physical ground state is \(\cos(2\phi)\), which corresponds to the following local operator in the UV: 
\begin{equation}
\label{Eq:order_hh}
\tilde{\mathcal{Q}}^{}_j = (Z_j Z_{j+1} - X_j X_{j+1}) + (j\rightarrow j+1).
\end{equation}
The fixed-point Hamiltonian takes the form
\begin{equation}
\label{Eq:hh_breaking}
\tilde{H} = \sum_j (I- X_j X_{j+1})( I - Z_{j+2} Z_{j+3}) + (Z \leftrightarrow X).
\end{equation}
Similar to the VBS and the $(\eta^{Z}, \eta^{X})$-breaking phases, the fixed-point Hamiltonian cannot be written as a sum of commuting projectors. However, one can verify that $\tilde{H}$ has a finite gap with aforementioned wave functions as its only ground states.
In fact, the fixed-point Hamiltonian of $(\eta^{\mathsf{H}}, \eta^{\mathsf{H^-}})$-breaking phase is related to the $(\eta^{Z}, \eta^{X})$-breaking phase  by the product of single-site unitary $U^Y(\theta ) = e^{-i \theta \sum_j Y_j/2}$ with $\theta = \pi/4$. 
We note this doesn't contradict with the fact that $(\eta^{Z}, \eta^{X})$-breaking and  $(\eta^{\mathsf{H}}, \eta^{\mathsf{H^-}})$-breaking phases are distinct phases of matter, as $U^Y(\theta = \pi/4)$ is a short-depth unitary that breaks the symmetry generators $\eta^Z, \eta^X,$ and $\eta^{\mathsf{H}}$.

Now, the original Hamiltonian can be obtained by undoing the Baxter transformation to Eq\eqref{Eq:hh_breaking}, and one finds
\begin{equation}
\label{Eq:SandD_fixedpoint_ham}
\begin{aligned}
    H = \sum_{j}   (I - & Z^e_{j}Z^e_{j+1} ) (I- Z^o_{j+1}Z^o_{j+2}) \\
    & +(I -  X^o_{j}) (I- X^e_{j+1})+ (o \leftrightarrow e).
\end{aligned}
\end{equation}
One can easily see that the ground-state subspace includes $|x,+\rangle_o |z,\pm\rangle_e$, $|z,\pm \rangle_o |x,+\rangle_e$, as they are all eigenvectors of $H$ with eigenvalue $0$. We call this phase the $(S, D^o D^e)$-breaking phase as all physical ground states have vanishing expectation values with respect to the swap and diagonal Kramers-Wannier symmetry operators.
{The four-fold degeneracy can also be derived using the topological properties of the Baxter transformation, and the idea is similar to the ones we have done in $(D^o D^e, S D^o D^e)$-breaking and symmetric phases.
We now provide a more illuminating way to see the four-fold degeneracy, which also makes connection to the system with symmetry generators ($\eta^o, \eta^e$, $S D^o D^e$) in Fig.\ref{fig:main_results}(a).
If one considers only ($\eta^o, \eta^e$, $S D^o D^e$) as symmetry generators, the most relevant perturbation involving the fields $\phi$ is $\cos(2 \phi)$, which drives the gauged system to $|z,\pm\rangle$ (corresponds to $|z,\pm \rangle_o |x,+\rangle_e$ in the original system) and $|x,\pm\rangle$ (corresponds to $|x,+\rangle_o |z,\pm\rangle_e$ in the original system)  depending on the sign of the interaction. 
Therefore, the $(S, D^o D^e)$-breaking phase, while being a stable phase with the symmetry generators ($\eta^o, \eta^e$, $S D^o D^e$, $S$), has now become the first-order line between the $\eta^o$-breaking (with fixed-point wavefunctions $|z,\pm \rangle_o |x,+\rangle_e$) and $\eta^e$-breaking (with fixed-point wavefunctions $|x,+\rangle_o |z,\pm\rangle_e$) phases in the absence of the swap symmetry. 
Since the fixed point Hamiltonians for both $\eta^o$-breaking and $\eta^e$-breaking phases can be written as sums of commuting-projectors, the four-fold degeneracy in the original model is apparent from the coexistence of the two phases.
Finally, we note as a side remark that the existence of the $\eta^o$-breaking and $\eta^e$-breaking phases under the symmetry generators $(\eta^o, \eta^e, S D^o D^e)$ is consistent with the Rep$(D_8)$ fusion category discussed in Ref.~\cite{thorngren2024fusion}, where the partition functions of both phases are invariant under gauging the $\eta^o \eta^e$ symmetry with off-diagonal pairing.
}

We now identify the local order parameter that characterizes the $(S, D^o D^e)$-breaking phase. Similar to the $(D^o D^e, S D^o D^e)$-breaking phase, since $\tilde{\mathcal{Q}}$ in Eq.~\eqref{Eq:order_hh} transforms trivially under $\eta^X$ and $\eta^Z$, it remains local after reversing the Baxter transformation and takes the form
\begin{equation}
\mathcal{Q}_{j} = (X^o_j - Z^o_j Z^o_{j+1}) - (o \leftrightarrow e).
\end{equation}
As with $\mathcal{O}_j$, $\mathcal{Q}_j$ does not transform simply under the $D^o D^e$ symmetry. However, in the low-energy limit, $\mathcal{Q}_j$ carries a charge under $D^o D^e$, since any low-energy state $|\psi\rangle$ in the neutral charge sector satisfies $(D^o D^e)^2 |\psi\rangle = |\psi\rangle$. One can similarly show that $\mathcal{Q}_j$ is charged under $S$ but remains invariant under $S D^o D^e$. These properties establish $\mathcal{Q}_j$ as a local order parameter that distinguishes the $(S, D^o D^e)$-breaking phase from other phases.

\end{itemize}

\subsection{Phase transitions}

We now consider phase transitions among those phases described by the action in Eq.\eqref{Eq:field_SDoDe}, i.e., $S_{\text{int} }= \int d\tau d x  \big[\lambda_\theta \cos(2\theta) +\lambda_\phi \cos(4 \phi)  \big]$. The scaling dimensions of the interactions close to the Gaussian fixed point are:
\begin{equation}
\label{Eq:scaling_dimension}
 \text{dim}[\cos(n \theta)] = {n^2 g},\ \text{dim}[\cos(m \phi)] = \frac{m^2}{4g}.
\end{equation}
Eq.\eqref{Eq:scaling_dimension} implies  $\text{dim}[\cos(2 \theta)] =4g$ and $\text{dim}[\cos(4 \phi)] = 4/g$. Therefore, when $g = 1/2 $ ($g= 2$), $\text{dim}[\cos(4 \theta)](\text{dim}[\cos(4 \phi)])$ equals two, the spacetime dimension.
It is then natural to consider three separated cases: $g<1/2$, $g>2$, and $2>g>1/2$.
When $g<1/2$, the $\lambda_\theta (\lambda_\phi)$ term is relevant(irrelevant), and thus a continuous transition occurs when the sign of the single parameter $\lambda_{\theta}$ changes. This corresponds to a phase transition between the partially ordered and $(D^o D^e, S D^o D^e)$-breaking phases. 
Similarly, when $g>2$, the $\lambda_\phi (\lambda_\theta)$ term is relevant(irrelevant), and thus a continuous transition occurs when the sign of the single parameter $\lambda_{\phi}$ changes. This corresponds to a phase transition between the symmetric and $(S, D^o D^e)$-breaking phases.
On the other hand, when $2>g>1/2$, both $\lambda_\phi$ and $\lambda_\theta$ are irrelevant, and thus the system remains gapless. Since this regime separates the $|\lambda_{\theta}|$-large and $|\lambda
_{\phi}|$-large phases, there is no direct transition between the $(D^o D^e, S D^o D^e)$-breaking and $(S, D^o D^e)$-breaking phases [see Fig.\ref{fig:main_results}(b)].

{
We now briefly discuss phases and phase transitions when only $(\eta^o, \eta^e, S D^o D^e)$ are considered as symmetry generators. In this situation, the gauged model respects the symmetry generators ($\eta^X, \eta^Z$, $T$), and hence the most relevant symmetry-preserving interaction takes the form 
\begin{equation}
\label{Eq:singe_copy_Sint}
S_{\text{int} }= \int d\tau d x  \big[\lambda_\theta \cos(2\theta) +\lambda'_\phi \cos(2 \phi)\big].
\end{equation}
As briefly mentioned in Sec.\ref{sec:identify_phases}, $\lambda'_\phi \rightarrow  \infty$ and $\lambda'_\phi \rightarrow  -\infty$ correspond to $x$-FM (the $\eta^e$-breaking phase in the original model) and $z$-FM (the $\eta^o$-breaking phase in the original model) phases respectively. 
{
We note that the $\eta^o$- and $\eta^e$-breaking phases cannot exist individually once the swap symmetry is enforced, as the swap exchanges them. The $(S, D^o D^e)$-breaking phase discussed in Sec.~\ref{sec:identify_phases} originates from these two phases.
}
On the other hand, $\lambda_\theta \rightarrow \pm \infty$ still drives the system to partially ordered and $(D^o D^e, S D^o D^e)$-breaking phases.
Phase transitions among those phases can be analyzed by nothing that
Eq.\eqref{Eq:scaling_dimension} implies  $\text{dim}[\cos(2 \theta)] =4g$ and $\text{dim}[\cos(2 \phi)] = 1/g$. Therefore, when $g = 1/2 $ both $\cos(4 \theta)$ and $\cos(2 \phi)$ are marginal.
The situation for $g<1/2$ behaves exactly the same as the previous situation.
On the other hand, when $g>1/2$, a continuous transition between $\eta^e$-breaking and $\eta^o$-breaking phases can be tuned by a single parameter $\lambda'_\phi$ [see Fig.\ref{fig:main_results}(a)].
This implies that the originally fine-tuned transition between two ordinary symmetry breaking phases has now become generic due to the presence of the non-invertible symmetry $S D^o D^e$.
Finally, another interesting situation arises when $g \approx 1/2$, where a transition between the $|\lambda_\theta|$-large and $|\lambda'_\phi|$-large phases is expected to be tunable by a single parameter $g$, with a critical point near $g_c \approx 1$. While the critical theory contains finely balanced cosine terms in the current field variables, Ref.~\cite{jiang2019ising} has shown that, upon rewriting the theory in an alternative set of variables, it can be mapped to a Gaussian theory with only one relevant term. Therefore, a single-parameter-tuned continuous transition between the $|\lambda_\theta|$-large and $|\lambda'_\phi|$-large phases exists. We will see in the next section that a similar scenario occurs when considering two coupled $\mathbb{Z}^o_2 \times \mathbb{Z}^e_2$ spin chains.
}

\section{Two-coupled spin chains}
\label{sec:two_copy}

In the previous section, we find that the $(S, D^o D^e)$-breaking and $(D^o D^e, S D^o D^e)$-breaking phases are separated by a gapless regime [see Fig.\ref{fig:main_results}(b)].
To search for a direct second-order transition between the two non-invertible SSB phases, we consider two copies of $\mathbb{Z}^o_2 \times \mathbb{Z}^e_2$ spin chains with $\prod_{r = 1}^2 S^{(r)}  $  and $\prod_{r = 1}^2 S^{(r)} D^{o,r} D^{e,r}$ symmetries. We further require the system remains invariant under the exchange of the two copies (i.e., $1\leftrightarrow 2$), which we denote as $\mathcal{E}$. 
We will demonstrate that interactions between the two copies can generate relevant perturbations that destabilize the critical regime present in the decoupled limit.

To make connection with the field theories in previous sections, we apply the Baxter transformation on each copy:
\begin{equation}
    \begin{aligned}
        X^{o, (r)}_{j}  & \rightsquigarrow  X^{(r)}_{2j-1} X^{(r)}_{2j} ,  \\
        X^{e, (r)}_{j}  & \rightsquigarrow  Z^{(r)}_{2j-1} Z^{(r)}_{2j} ,  \\
         Z^{o, (r)}_{j}Z^{o, (r)}_{ j+1} &\rightsquigarrow  Z^{(r)}_{2j} Z^{(r)}_{2j+1} ,  \\
         Z^{e, (r)}_{j}Z^{e, (r)}_{j+1} & \rightsquigarrow  X^{(r)}_{2j} X^{(r)}_{2j+1}, \\
    \end{aligned}
\end{equation}
where $r = 1,2$ denotes the $r$-th copy.
The gauged model is then invariant under the symmetry generators $\eta^{Z,1}, \eta^{X,1}, \eta^{Z,2}, \eta^{X,2}$, $\prod_{r = 1}^{2}\eta^{\mathsf{H}, r} $, $\prod_{r = 1}^{2} T^{(r)}$, and ${\mathcal{E}}$. 
Therefore, we consider the following Hamiltonian
\begin{equation}
\begin{aligned}
    \tilde{H} =  & -\sum_{j=1}^{2L}  [X^{(1)}_j X^{(1)}_{j+1}+ Z^{(1)}_j Z^{(1)}_{j+1} + (1 \leftrightarrow 2)]   + \delta \tilde{H}_{\text{int}},
\end{aligned}
\end{equation}
where $\delta \tilde{H}_{\text{int}}$ denotes all possible symmetry-allowed perturbations. Our main result is shown in Fig.~\ref{fig:DQCP}, where we find that there exists a second-order transition tuned by a single parameter between the analogous non-invertible SSB phases discussed in Sec.~\ref{sec:second_scenario}.
Interestingly, in the vicinity of the critical surface (including the gapped regime close to it), there is an emergent non-invertible ``cosine" symmetry originating from the emergent $U(1)$ symmetry when the theory is rewritten using a suitable set of field variables. 
We note that similar non-invertible continuous symmetries have been discussed in Refs.~\cite{chang2021lorentzian, thorngren2024fusion, eck2024from, cao2025global, seifnashri2025gauging} in the context of gauging the charge conjugation symmetry of the compact free boson.
{We conjecture that this emergent symmetry leads to an emergent anomaly, thereby forbidding the existence of a symmetric gapped phase close to the critical surface. }
Furthermore, the critical surface exhibits an enhanced Kramers--Wannier-like self-duality, characterized by the additional generators $D^{\alpha,1} D^{\alpha,2}$, where $\alpha = o,e$. This enhanced symmetry exchanges the order parameters between the two non-invertible SSB phases (up to a sign), demonstrating that the two order parameters have the same scaling dimension at the critical points.
Finally, the DQCP we identified persists as DQCP under any gauging scheme involving the spin-flip symmetries. We will demonstrate how to use this robustness to generate a large family of related DQCPs.

Similar to the previous case, the way the symmetry generators in the gauged model act in the low-energy limit can be identified using bosonization, and one finds
\begin{equation}
\label{Eq:twocopy_lowenergy}
    \begin{aligned}
    \eta^{Z,1}:\ & (\phi^{(1)}, \theta^{(1)} ) \rightarrow (-\phi^{(1)}, -\theta^{(1)} ), \\   
    \eta^{X,1}:\ & (\phi^{(1)}, \theta^{(1)} ) \rightarrow (-\phi^{(1)} + \pi, -\theta^{(1)}), \\    
    {\eta}^{Z,2}:\ & ({\phi}^{(2)}, {\theta}^{(2)} ) \rightarrow (-{\phi}^{(2)}, -{\theta}^{(2)} ), \\   
    {\eta}^{X,2}:\ & ({\phi}^{(2)}, {\theta}^{(2)} ) \rightarrow (-{\phi}^{(2)} + \pi, -{\theta}^{(2)}), \\    
    \prod_{r = 1}^{2}{\eta}^{\mathsf{H}, r}:\ & (\phi^{(1)}, \theta^{(1)}, {\phi}^{(2)}, {\theta}^{(2)} ) \\
    & \rightarrow (-\phi^{(1)} - \frac{\pi}{2}, -\theta^{(1)}, -{\phi}^{(2)} - \frac{\pi}{2}, -{\theta}^{(2)}).\\
   \prod_{r = 1}^{2}  T^{(r)} :\ & (\phi^{(1)}, \theta^{(1)}, {\phi}^{(2)},  \bar{\theta}^{(2)}) \\
    & \rightarrow (\phi^{(1)}, \theta^{(1)} + {\pi}, {\phi}^{(2)},  {\theta}^{(2)} + {\pi}), \\
    \mathcal{E}:\ &  (\phi^{(1)}, \theta^{(1)}, {\phi}^{(2)}, {\theta}^{(2)} ) \rightarrow  (\phi^{(2)}, \theta^{(2)}, {\phi}^{(1)}, {\theta}^{(1)} ) \\
    \end{aligned}
\end{equation}
The symmetry-preserving action can then be written as
\begin{equation}
\label{Eq:action_two_copy}
S = \big(\sum_{r = 1,2} S_0^{(r)}[\phi^{(r)}, \theta^{(r)}] \big)+ S_{\text{int}}[\phi^{(1)}, \theta^{(1)}, \phi^{(2)}, \theta^{(2)}]
\end{equation}
Here, $S_0^{(1)}$ and $S_0^{(2)}$ take the same form as in Eq.\eqref{Eq:S_0}, i.e. $S^{(r)}_0 = \int d\tau d x   \Big[ \frac{i}{2 \pi} \partial_\tau \phi^{(r)}  \partial_x \theta^{(r)} + \frac{v}{2\pi} \Big(\frac{1}{4 g}(\partial_x \theta^{(r)})^2 + g (\partial_x \phi^{(r)})^2 \Big) \Big], 
$ $r = 1,2$.
On the other hand, the most relevant terms in $S_{\text{int}}$ can be easily identified by noting that we seek inter-chain interactions that are forbidden in the presence of individual $\eta^{\mathsf{H},r}$ and $T^{(r)}$ symmetries but now become allowed due to the weakened diagonal symmetries $\eta^{\mathsf{H},1} \eta^{\mathsf{H},2}$ and $T^{(1)} T^{(2)}$. In particular, $S_{\text{int}}$ takes the form
\begin{equation}
\begin{aligned}
S_{\text{int}} = \int d\tau dx \Big\{& \Lambda_\theta \cos(\theta^{(1)}) \cos(\theta^{(2)}) \\
& + \Lambda_\phi \cos(2\phi^{(1)}) \cos(2\phi^{(2)}) + \cdots \Big\},
\end{aligned}
\end{equation}
where $\cdots$ denotes less relevant terms.
Since we are interested in the regime $2 > g > 1/2$, where each spin chain exhibits a stable critical regime when decoupled, we do not include interactions that couple fields within each individual chain, as such interactions are irrelevant in this regime.

Similar to the single-copy case, all gapped phases can be identified by considering the limits $\Lambda_\theta \rightarrow \pm \infty$ and $\Lambda_\phi \rightarrow \pm \infty$. 
In particular, the limit $\Lambda_\theta \rightarrow -\infty$ forces the fields to satisfy $\cos(\theta^{(1)}) = \cos(\theta^{(2)}) = \pm 1$. The fixed-point wave functions of the original model include $|x,+ \rangle_{o,1}  |x,+ \rangle_{e,1} \otimes |(1 \leftrightarrow 2)\rangle$ and $|z,\pm \rangle_{o,1}  |z,\pm \rangle_{e,1} \otimes |(1 \leftrightarrow 2)\rangle$, and thus the ground-state subspace is 17-dimensional (since $17 = 1 + 2^4$). This phase is analogous to the $(D^o D^e, S D^o D^e)$-breaking phase discussed in Sec.~\ref{sec:second_scenario}, since the second copy is identical to the first (i.e., the system does not break the exchange $\mathcal{E}$ symmetry).
On the other hand, $\Lambda_\phi \rightarrow -\infty$ imposes $\cos(2\phi^{(1)}) = \cos(2\phi^{(2)}) = \pm 1$. The fixed-point wave functions of the original model include $|x,+ \rangle_{o,1}  |z,\pm \rangle_{e,1}  \otimes |(1 \leftrightarrow 2)\rangle$ and $|z,\pm \rangle_{o,1}  |x,+ \rangle_{e,1} \otimes |(1 \leftrightarrow 2)\rangle$, and thus the ground-state subspace is 8-dimensional (since $8 = 2^2 + 2^2$). This phase is analogous to the $(S, D^o D^e)$-breaking phase discussed in Sec.~\ref{sec:second_scenario}.
One can also derive the fixed-point wave functions in both $\Lambda_\theta \rightarrow +\infty$ and $\Lambda_\phi \rightarrow +\infty$ limits. 
For simplicity, we will focus on the cases where $\Lambda_\theta = -|\Lambda_\theta|$ and $\Lambda_\phi = -|\Lambda_\phi|$, and we refer to them as the $(\prod_{r = 1}^2 D^{o,r} D^{e,r}, \prod_{r = 1}^2 S^{(r)} D^{o,r} D^{e,r})$-breaking and $(\prod_{r = 1}^2 S^{(r)}, \prod_{r = 1}^2 D^{o,r} D^{e,r})$-breaking phases. 

The corresponding order parameters for $(\prod_{r = 1}^2 D^{o,r} D^{e,r}, \prod_{r = 1}^2 S^{(r)} D^{o,r} D^{e,r})$-breaking and $(\prod_{r = 1}^2 S^{(r)}, \prod_{r = 1}^2  D^{o,r} D^{e,r})$-breaking phases can similarly be identified, as in Sec.~\ref{sec:second_scenario}, as $\mathcal{O}_j = (X^{o,(1)}_j - Z^{o,(1)}_j Z^{o,(1)}_j) + (o \leftrightarrow e) \sim \cos(\theta^{(1)})$ and $\mathcal{Q}_j = (X^{o,(1)}_j - Z^{o,(1)}_j Z^{o,(1)}_j) - (o \leftrightarrow e) \sim \cos(2\phi^{(1)})$, respectively (since both phases preserve the symmetry $\mathcal{E}$ that exchanges the first and the second copy, it suffices to consider the local order parameters that only involve the operators in the first copy).
Before analyzing the phase transition between the two phases, we present the fixed-point lattice Hamiltonians for both, for concreteness. For the $(\prod_{r = 1}^2 D^{o,r} D^{e,r}, \prod_{r = 1}^2 S^{(r)} D^{o,r} D^{e,r})$-breaking phase, the gauged model Hamiltonian is given by $\tilde{H} = \sum_j \sum_{r,s=1}^2 h^{r,s}_{j,j+1}$, where
\begin{equation}
h^{r,s}_{j,j+1} = (I - Z^{(r)}_j Z^{(r)}_{j+1})(I - Z^{(s)}_{j+1} Z^{(s)}_{j+2}) + (Z \leftrightarrow X).
\end{equation}
For the $(\prod_{r = 1}^2 S^{(r)}, \prod_{r = 1}^2 D^{o,r} D^{e,r})$-breaking phase, the gauged model Hamiltonian also takes the form $\tilde{H} = \sum_j \sum_{r,s=1}^2 h^{r,s}_{j,j+1}$, with
\begin{equation}
h^{r,s}_{j,j+1} = (I - X^{(r)}_j X^{(r)}_{j+1})(I - Z^{(s)}_{j+2} Z^{(s)}_{j+3}) + (Z \leftrightarrow X).
\end{equation}
The Hamiltonians in the original model can be obtained straightforwardly by reversing the Baxter transformation.

Let's now consider the phase transition between the $(\prod_{r = 1}^2 D^{o,r} D^{e,r}, \prod_{r = 1}^2 S^{(r)} D^{o,r} D^{e,r})$-breaking and $(\prod_{r = 1}^2 S^{(r)}, \prod_{r = 1}^2  D^{o,r} D^{e,r})$-breaking phases.
Using Eq.\eqref{Eq:scaling_dimension}, one finds $\text{dim}[\cos( \theta) \cos(\bar{\theta} )] =2g$ and $\text{dim}[\cos(2 \phi) \cos(2 \bar{\phi})] =2/g$.
Therefore, when $g<1 (g>1)$, $\cos(\theta) \cos( \bar{\theta} ) [\cos(2 \phi) \cos(2 \bar{\phi})  ]$ is relevant, and hence the system is in the $(\prod_{r = 1}^2  D^{o,r} D^{e,r}, \prod_{r = 1}^2 S^{(r)} D^{o,r} D^{e,r})$-breaking [$(\prod_{r = 1}^2 S^{(r)},\prod_{r = 1}^2 D^{o,r} D^{e,r})$-breaking] phase.
On the other hand, when $g = 1$ both perturbations are marginal. A transition between two non-invertible SSB phases is then expected to be tuned by a single parameter $g$ with the transition point $g_c \approx 1$.
However, the critical theory that describes this transition contains finely balanced cosine terms expressed in the current field variables, resulting in a non-perturbative situation. 
In the following, we will show that by rewriting the theory in an alternative set of field variables, the critical theory in the regime $|\Lambda_\theta| + |\Lambda_\phi| \gg ||\Lambda_\theta| - |\Lambda_\phi||, |g-1|$ can be mapped to a Gaussian theory with only one relevant term. 
We note that this situation is similar to the $(1+1)$-D DQCP discussed in Ref.~\cite{jiang2019ising}, which studies the transition from an Ising ferromagnet to a valence bond solid. 
Furthermore, it will become evident from the rewritten variables that the theory close to the critical surface possesses an emergent $U(1)$ symmetry, which corresponds to a non-invertible symmetry in the original model.

\subsection{Critical theory in alternative set of field variables}
\label{sec:alternative}

\begin{figure}
\centering
\includegraphics[width=\linewidth]{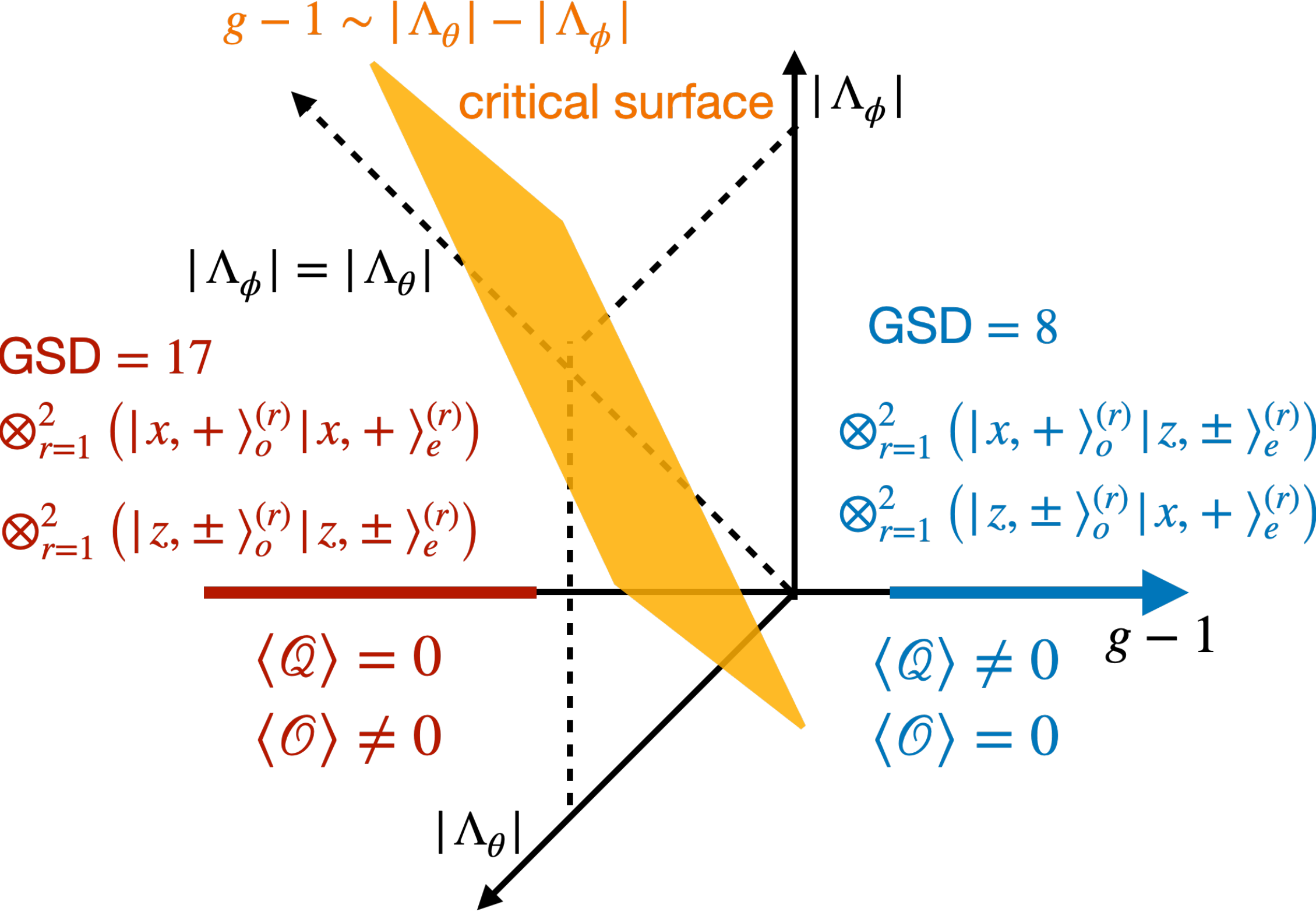}
\caption{ 
Phase diagram of Eq.\eqref{Eq:action_two_copy} in the regime $|\Lambda_\theta| + |\Lambda_\phi| \gg ||\Lambda_\theta| - |\Lambda_\phi||, |g-1|$ . There exists a critical surface between the $(\prod_{r = 1}^2 D^{o,r} D^{e,r}, \prod_{r = 1}^2 S^{(r)} D^{o,r} D^{e,r})$-breaking and $(\prod_{r = 1}^2 S^{(r)}, \prod_{r = 1}^2  D^{o,r} D^{e,r})$-breaking phases, and thus a second-order transition can be tuned by a single parameter.  The theory at the critical surface is described by a single Luttinger parameter $K = 1- \frac{2 \pi}{v} (|\Lambda_\phi| + |\Lambda_\theta|)<1$ and the scaling dimensions of the order parameters in both phases are equivalent to the Luttinger parameter $K$.
}
\label{fig:DQCP}
\end{figure}

The first step of rewriting the theory is to map the coupled bosonic fields $(\phi^{(1)}, \theta^{(1)}, \phi^{(2)}, \theta^{(2)})$ into eight coupled Majorana fields $(\gamma^o_{L/R}, \gamma^e_{L/R}, \eta^o_{L/R}, \eta^e_{L/R})$, This is done by identifying $\gamma^o_R + i \gamma^e_R \sim e^{i(\phi^{(1)} + \theta^{(1)}/2)}$, $\gamma^o_L + i \gamma^e_L \sim e^{-i(\phi^{(1)} - \theta^{(1)}/2)}$, and $ \eta^o_R + i \eta^e_R \sim e^{i(\phi^{(2)} + \theta^{(2)}/2)}$, $ \eta^o_L + i \eta^e_L \sim e^{-i(\phi^{(2)} - \theta^{(2)}/2)}$.
The Lagrangian can then be written as $\mathcal{L} = \mathcal{L}_1 + \mathcal{L}_{\text{int}}$, where $\mathcal{L}_1$ is a sum of eight decoupled free Majorana fields. On the other hand, $\mathcal{L}_{\text{int}} $ can be further decomposed as
$\mathcal{L}_{\text{int}} = \mathcal{L}_{g-1} +\mathcal{L}_{\phi} + \mathcal{L}_\theta $, where
\begin{equation}
\begin{aligned}
\mathcal{L}_{g-1} & \sim (g-1) (\partial_x \phi)^2 \sim -(g-1)   (\gamma^o_L \gamma^o_R \gamma^e_L \gamma^e_R+ \eta^o_L \eta^o_R \eta^e_L \eta^e_R), \\
\mathcal{L}_{\phi} & \sim |\Lambda_\phi| (\gamma^o_L \gamma^o_R + \gamma^e_L \gamma^e_R)(\eta^o_L \eta^o_R +\eta^e_L \eta^e_R), \\ 
\mathcal{L}_{\theta}& \sim |\Lambda_\theta| (\gamma^o_L \gamma^o_R - \gamma^e_L \gamma^e_R)(\eta^o_L \eta^o_R -\eta^o_L \eta^o_R).
\end{aligned}
\end{equation}
Now, we re-pair the four Majorana fields $(\gamma^o_{L/R}, \eta^o_{L/R})$ into new bosonic fields $(\phi^{(o)}, \theta^{(o)})$, and similarly, pair $(\gamma^e_{L/R}, \eta^e_{L/R})$ into new bosonic fields $(\phi^{(e)}, \theta^{(e)})$. Specifically, we identify
$\gamma^{o/e}_R + i \eta^{o/e}_R \sim e^{i({\phi}^{o/e} + {\theta}^{o/e}/2)}$ and $ \gamma^{o/e}_L + i \gamma^{o/e}_L \sim e^{-i({\phi}^{o/e} - {\theta}^{o/e}/2)}$. 
It follows that $\mathcal{L} = \mathcal{L}_1+\mathcal{L}_{\text{int}}$, where $\mathcal{L}_1$ is Lagrangian of the decoupled compact free bosons:
\begin{equation}
 \mathcal{L}_1 = \sum_{\alpha = o,e} \frac{i}{2\pi} \partial_\tau \phi^{\alpha}  \partial_x \theta^{\alpha} + \frac{v}{2\pi} \Big(\frac{1}{4 K}(\partial_x \theta^{\alpha})^2 + K (\partial_x \phi^{\alpha} )^2 \Big),
\end{equation}
with the Luttinger parameter $K = 1$. On the other hand, the interaction term can be further decomposed as
$\mathcal{L}_{\text{int}} = \mathcal{L}_{\phi+\theta} + \mathcal{L}_{\phi-\theta} +  \mathcal{L}_{g-1} $, where
\begin{equation}
\begin{aligned}
\mathcal{L}_{\phi+ \theta} & \sim ( |\Lambda_\phi| + |\Lambda_\theta| ) (\gamma^o_L \gamma^o_R \eta^o_L \eta^o_R+ \eta^e_L \eta^e_R \eta^e_L \eta^e_R) \\
& \sim -(|\Lambda_\phi| + |\Lambda_\theta|) [(\partial_x \phi^{o})^2 + (\partial_x \phi^{e})^2] , \\
\mathcal{L}_{\phi - \theta} & \sim (|\Lambda_\phi| - |\Lambda_\theta| )(\gamma^o_L \gamma^o_R \eta^e_L \eta^e_R+ \eta^o_L \eta^o_R \gamma^e_L \gamma^e_R ) \\
& \sim -{|\Lambda_\phi| - |\Lambda_\theta|} [\cos(2 \phi^o)\cos(2 \phi^e) - \cos( \theta^o)\cos( \theta^e)], \\ 
\mathcal{L}_{g-1} & \sim {(g-1)} [\cos(2 \phi^o)\cos(2 \phi^e) + \cos(\theta^o)\cos( \theta^e)].
\end{aligned}
\end{equation}
Therefore, we find that the effect of $\mathcal{L}_{\phi + \theta}$ shifts the Luttinger parameter from $K = 1$ to $K = 1- \frac{2 \pi}{v} (|\Lambda_\phi| + |\Lambda_\theta|)$ with $v$ a positive non-universal parameter.
In this regime, the $\cos(2\phi^o)\cos(2\phi^e)$-term is irrelevant (since $\text{dim}[\cos(2\phi^o)\cos(2\phi^e) ]= 2/K >2$) and thus vanishes in the infrared limit. The remaining interaction takes the form
\begin{equation}
\label{Eq:rewrtie_Lint}
\mathcal{L}_{\text{int}} \sim -{(|\Lambda_\theta| - |\Lambda_\phi| - \delta g)}\cos(\theta^{(o)})\cos( \theta^{(e)}),
\end{equation}
where $\delta g $ is proportional $ g-1$ up to a non-universal constant.
Therefore, the theory in the regime $|\Lambda_\theta| + |\Lambda_\phi| \gg ||\Lambda_\theta| - |\Lambda_\phi||, |g-1|$ is described by a single relevant cosine term, and a continuous transition occurs when the the coefficient $(|\Lambda_\theta| - |\Lambda_\phi| - \delta g)$ vanishes. It follows that the second-order transition between the $(\prod_{r = 1}^2 D^{o,r} D^{e,r}, \prod_{r = 1}^2 S^{(r)} D^{o,r} D^{e,r})$-breaking and $(\prod_{r = 1}^2 S^{(r)}, \prod_{r = 1}^2  D^{o,r} D^{e,r})$-breaking phases can be tuned by a single parameter $g$.  
The scaling dimensions of the order parameters in both phases are determined by the Luttinger parameter $K = 1 - \frac{2\pi}{v} (|\Lambda_\phi| + |\Lambda_\theta|)$, and satisfy $\text{dim}[\mathcal{O}] = \text{dim}[\mathcal{Q}] = K$.

\subsubsection*{Emergent cosine symmetry close to the critical surface.}

Interestingly, the fact that the $\cos(2\phi^\alpha)\cos(2\phi^\beta)$-term with $\alpha, \beta = o,e$ being irrelevant implies that there is an emergent $U(1)$ symmetry $\phi^{\alpha} \rightarrow \phi^\alpha + a$, $a \in [0,2\pi)$ in the current set of variables. We now show that this corresponds to an additional non-invertible symmetry in the original model.
To see this, we note that the aforementioned rewriting of the theory can be done exactly on the lattice, and in fact, it corresponds to applying two Baxter transformations, one on the odd chains (of both copies) and one on the even chains. Specifically, denoting $\alpha = o,e$, we consider the following transformation:
\begin{equation}
    \begin{aligned}
        X^{\alpha,1}_{j}  & \rightsquigarrow  X^{\alpha}_{2j-1} X^{\alpha}_{2j} ,  \\
        X^{\alpha,2}_{j}  & \rightsquigarrow  Z^{\alpha}_{2j-1} Z^{\alpha}_{2j} ,  \\
         Z^{\alpha, 1}_{j}Z^{\alpha, 1}_{ j+1} &\rightsquigarrow  Z^{\alpha}_{2j} Z^{\alpha}_{2j+1} ,  \\
         Z^{\alpha, 2}_{j}Z^{\alpha, 2}_{j+1} & \rightsquigarrow  X^\alpha_{2j} X^\alpha_{2j+1}. \\
    \end{aligned}
\end{equation}
At the Gaussian fixed point, these Pauli operators can be identified as $Z^\alpha_j \sim \cos(\phi^\alpha)$, $X^\alpha_j \sim -\sin(\phi^\alpha)$, and $Y^\alpha_j \sim  \frac{\partial_x \theta^\alpha}{\pi} + A(-1)^j \sin(\theta^\alpha)$.
Therefore, the emergent $U(1)$ symmetry is nothing but the rotation symmetry along the $y$-axis $U^\alpha_y(\lambda) = e^{i \lambda \tilde{Q}^\alpha}$ where the charge $\tilde{Q}^\alpha = (\sum_{j = 1}^{2L} Y^\alpha_{j})/2$. 
{
To translate this emergent $U(1)$ symmetry back to the original model, we note that since all operators charged under $\mathbb{Z}_2^Z \times \mathbb{Z}_2^X$ are projected out by the Baxter transformation, it suffices to consider the “cosine” operator $\cos(\lambda \tilde{Q}^\alpha) = \frac{1}{2}(e^{i \lambda \tilde{Q}^\alpha} + e^{-i \lambda \tilde{Q}^\alpha}) = \sum_{n=0}^\infty (-1)^n (\lambda \tilde{Q}^\alpha)^{2n}/(2n)!$, which is generated by powers of $(\tilde{Q}^\alpha)^2$. We note that $\cos(\lambda \tilde{Q}^\alpha)$ is non-invertible, as illustrated by the well-known multiplication formula $2\cos(a)\cos(b) = \cos(a+b) + \cos(a-b)$.
~\cite{chang2021lorentzian, thorngren2024fusion, eck2024from, cao2025global, seifnashri2025gauging}}. %
It is then straightforward to show that the Baxter transformation maps $(\tilde{Q}^\alpha)^2 $ to $(Q^\alpha)^2$, where
\begin{equation}
\begin{aligned}
(Q^\alpha)^2  & =  \frac{L^2}{2}  -  \frac{1}{2}\sum_{j=1}^{L-1} \sum_{k = j+1}^{L} \Bigg(  Z^{\alpha,1}_j Z^{\alpha,2}_j Z^{\alpha,1}_{k}  Z^{\alpha,2}_{k}  \times  \\ 
&  (I -X^{\alpha,1}_j X^{\alpha,2}_j)(I - X^{\alpha,1}_k X^{\alpha,2}_k) 
\prod_{l=j+1}^{k-1} (X^{\alpha,1}_l X^{\alpha,2}_l)\Bigg).
\end{aligned}
\end{equation}
{It is interesting to explore whether the combination of the cosine symmetry $\cos(\lambda {Q}^\alpha ) $ and the symmetry enforced in Eq.~\eqref{Eq:twocopy_lowenergy} forbids the existence of a symmetric gapped phase, thus leading to the emergent anomaly that underlies the DQCP we identify.}

\subsubsection*{Enhanced self-duality on the critical surface}
The explicit lattice realization of the rewriting of the theory also reveals the further enhanced symmetry on the critical surface. When the coefficient $(|\Lambda_\theta| - |\Lambda_\phi| - \delta g)$ vanishes, the theory acquires an extra symmetry $\eta^{\mathsf{H}, \alpha} T^{\alpha}: (\phi^\alpha, \theta^\alpha) \rightarrow (-\phi^\alpha - \pi/2, -\theta^\alpha + \pi)$, $\alpha = o,e$. By reversing the Baxter transformation, this corresponds to $D^{\alpha,1} D^{\alpha,2}$, the Kramers-Wannier transformation on the $\alpha$-th sites, in the original model. Since $D^{\alpha,1} D^{\alpha,2}$ exchanges the order parameters $\mathcal{O}$ and $\mathcal{Q}$ (up to a sign) in the low-energy limit, this enhanced symmetry provides one simple explanation why their scaling dimensions coincide.

\subsection{Generating new DQCPs through gauging}\label{sec:gauging}

\begin{figure*}
\centering
\includegraphics[width=\linewidth]{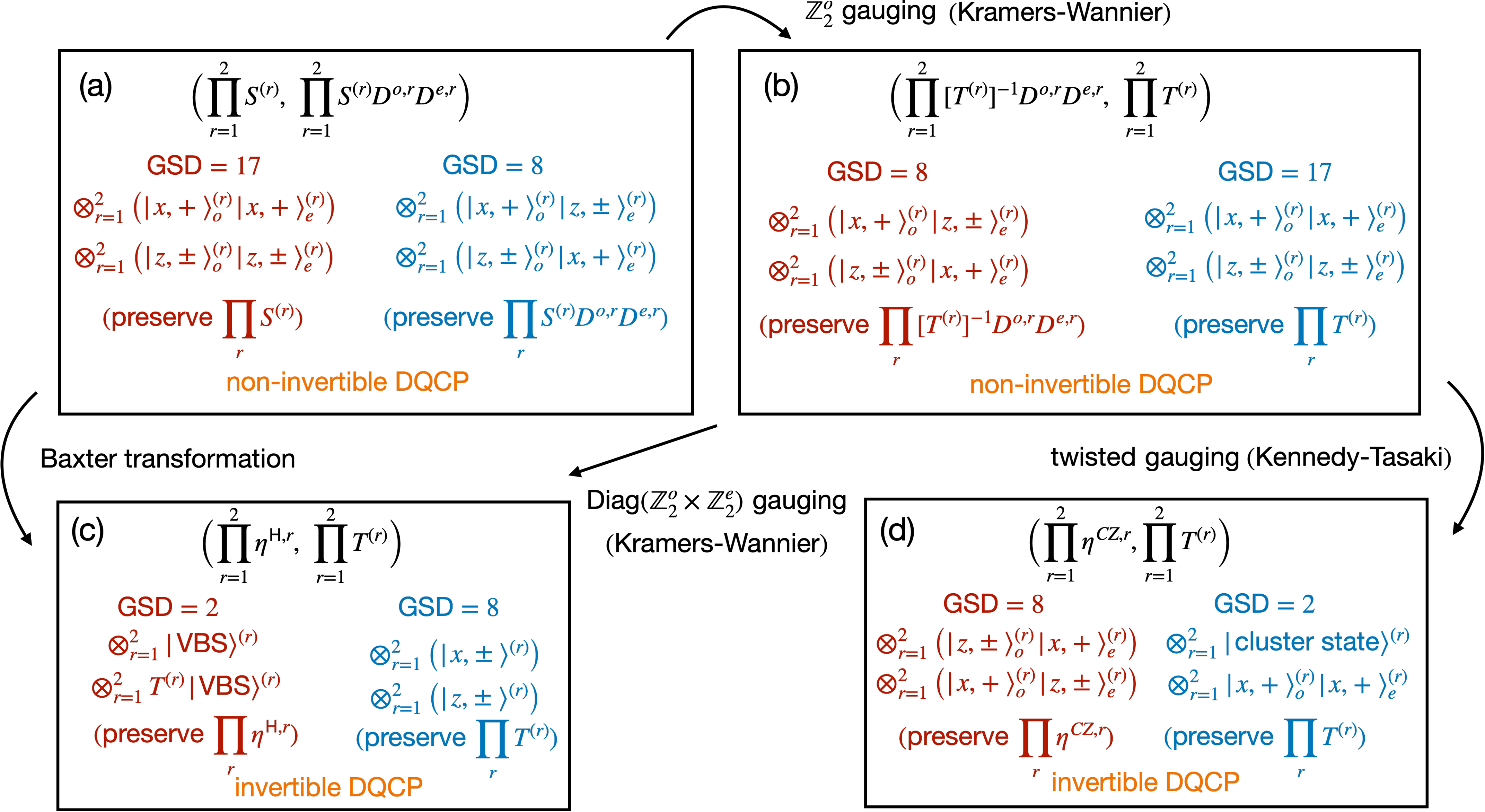}
\caption{New DQCPs from gauging.
Starting from the two-coupled $\mathbb{Z}^o_2 \times \mathbb{Z}^e_2$ symmetric spin chain that exhibits a non-invertible DQCP in (a), one can generate another non-invertible DQCP by gauging the $\mathbb{Z}_2^o$ symmetry on both copies, as shown in (b). From (b), one can obtain a new invertible DQCP with a type-III anomaly in (d) by applying twisted gauging (i.e., the Kennedy--Tasaki transformation). Alternatively, one can arrive at the previously studied invertible DQCP with an LSM anomaly in (c) using spin-chain bosonization by further gauging the diagonal $\mathbb{Z}_2^o \times \mathbb{Z}_2^e$ symmetry. Each subfigure lists the symmetry generators (excluding spin-flip and exchange symmetries), the ground-state degeneracies (GSDs) of each phase, the fixed-point wavefunctions, and the symmetries preserved in each phase of the corresponding DQCP.
}
\label{fig:newDQCP}
\end{figure*}

As mentioned in the introduction, an interesting consequence of a DQCP between two non-invertible SSB phases involving Kramers--Wannier symmetry is that it remains a DQCP under gauging the spin-flip symmetries. This follows from the fact that the order parameters characterizing the DQCP transform trivially under the spin-flip actions. The ``gauging-preserving" property of non-invertible DQCP has already manifested in the gauged model we studied: Recall that the Baxter transformation amounts to first gauging the $\mathbb{Z}_2^o$ symmetry, followed by gauging the diagonal subgroup $\tilde{\mathbb{Z}}_2^o \times \mathbb{Z}_2^e$, where $\tilde{\mathbb{Z}}_2^o$ is the dual of the $\mathbb{Z}_2^o$ symmetry. Therefore, a DQCP between the $(\prod_{r = 1}^2 D^{o,r} D^{e,r}, \prod_{r = 1}^2 S^{(r)} D^{o,r} D^{e,r})$-breaking and $(\prod_{r = 1}^2 S^{(r)}, \prod_{r = 1}^2  D^{o,r} D^{e,r})$-breaking phases in the original model implies that the gauged model also hosts a DQCP, now between the VBS and $(\prod_{r = 1}^2\eta^{\mathsf{H}, r}, \prod_{r = 1}^2\eta^{\mathsf{H}^-,r})$-breaking phases.
{
We emphasize that this is a distinctive property of the non-invertible DQCP, as gauging spin-flip symmetries in group-like DQCPs can sometimes result in an ordinary symmetry-breaking transition. As a concrete example, consider the invertible DQCP discussed in Fig.~\ref{fig:main_results}(a), which describes a transition between the $\eta^o$-breaking and $\eta^e$-breaking phases. If one gauges the $\mathbb{Z}_2^o$ symmetry, the $\eta^o$-breaking phase becomes symmetric (since the Kramers--Wannier transformation exchanges symmetric and symmetry-breaking phases), while the $\eta^e$-breaking phase remains unchanged.
}

In the following, we provide a few more examples to demonstrate the idea of using various gauging schemes to generate a large family of related DQCPs. See Fig.\ref{fig:newDQCP} for a summary.
The first example we consider involves gauging the $\mathbb{Z}_2^o$ symmetry of both copies in the original DQCP, under which the transformations $X^{o, (r)}_j \rightsquigarrow Z^{o, (r)}_j Z^{o, (r)}_{j+1}$ and $Z^{o, (r)}_j Z^{o, (r)}_{j+1} \rightsquigarrow X^{o, (r)}_{j+1}, \forall r = 1,2$ are applied.
One can easily show that, aside from the spin-flip symmetries and the symmetry $\mathcal{E}$ that exchanges the two copies, the resulting system possesses $\prod_{r = 1}^2[T^{(r)}]^{-1} D^{o,r} D^{e,r}$ (inherited from $\prod_{r = 1}^2 S^{(r)}$ in the original model) and $ \prod_{r = 1}^2 T^{(r)}$ (inherited from $\prod_{r = 1}^2 S^{(r)} D^{o,r} D^{e,r}$ in the original model) as symmetries. 
Here, $T^{(r)}$ is the translation operator on the $r$-th copy implementing $A^{(r)}_{j} \rightarrow A^{(r)}_{j+1}$ for any operator $A$ (recall $A^{o,(r)}_j \equiv A^{(r)}_{2j-1}$ and $A^{e,(r)}_j \equiv A^{(r)}_{2j}$ ). 
The corresponding local order parameters $\mathcal{O}'_j = (X^{e,(1)}_j -Z^{e,(1)}_j Z^{e,(1)}_{j+1} ) - ( o \leftrightarrow e) $ and $\mathcal{Q}'_j=(Z^{o,(1)}_j Z^{o,(1)}_{j+1} -X^{o,(1)}_j  ) + ( o \leftrightarrow e)$ are easily identified by applying the transformation to $\mathcal{O}_j$ and $\mathcal{Q}_j$, respectively. Notably, $\mathcal{O}'_j$ is charged under both $\prod_{r = 1}^2 D^{o,r} D^{e,r}$ and $\prod_{r = 1}^2 T^{(r)}$, while transforming trivially under $\prod_{r = 1}^2[T^{(r)}]^{-1} D^{o,r} D^{e,r}$. Conversely, $\mathcal{Q}'_j$ is charged under both $\prod_{r = 1}^2[T^{(r)}]^{-1} D^{o,r} D^{e,r}$ and $\prod_{r = 1}^2 D^{o,r} D^{e,r}$, while transforming trivially under $\prod_{r = 1}^2 T^{(r)}$. Therefore, the original non-invertible DQCP between the $(\prod_{r = 1}^2 D^{o,r} D^{e,r}, \prod_{r = 1}^2 S^{(r)} D^{o,r} D^{e,r})$-breaking and $(\prod_{r = 1}^2 S^{(r)} , \prod_{r = 1}^2 D^{o,r} D^{e,r})$-breaking phases is now mapped to another non-invertible DQCP between the $(\prod_{r = 1}^2  D^{o,r} D^{e,r}, \prod_{r = 1}^2 T^{(r)})$-breaking and $(\prod_{r = 1}^2[T^{(r)}]^{-1} D^{o,r} D^{e,r}, \prod_{r = 1}^2 D^{o,r} D^{e,r})$-breaking phases after gauging the $\mathbb{Z}_2^o$ symmetry of both copies.
The fixed-point wave functions for both phases also be easily identified.
The properties of this new non-invertible DQCP—including the symmetry generators (aside from the spin-flip and exchange symmetries), the ground-state degeneracies (GSDs) of each phase, the fixed-point wavefunctions, and the symmetries preserved by each phase—are summarized in the upper right box of Fig.~\ref{fig:newDQCP}.
As an aside, we note that one recovers the previously-studied gauged model with an intrinsic LSM anomaly if one further gauges the diagonal $\mathbb{Z}^{o}_2 \times \mathbb{Z}_2^e$ symmetries on both copies (under which the transformations $X^{(r)}_{j} \rightsquigarrow Z^{(r)}_{j} Z^{(r)}_{j+1}$ and $Z^{(r)}_{j} Z^{(r)}_{j+1} \rightsquigarrow X^{(r)}_{j+1}$ for all $r = 1,2$ are applied). We summarize the properties of this invertible DQCP in Fig.~\ref{fig:newDQCP}(c).

The second example we consider is to further apply twisted gauging of both copies —i.e., the Kennedy-Tasaki transformation~\cite{kennedy1992hidden,kennedy1992hiddenZ2}—to the first example. This corresponds to implementing the transformations $Z^{o,(r)}_j Z^{o,(r)}_{j+1} \rightsquigarrow Z^{o,(r)}_j X^{e,(r)}_j Z^{o,(r)}_{j+1}$ and $Z^{e,(r)}_j Z^{e,(r)}_{j+1} \rightsquigarrow Z^{e,(r)}_j X^{o,(r)}_{j+1} Z^{e,(r)}_{j+1}, \forall r = 1,2$. One can easily show that, aside from the spin-flip and the exchange $(\mathcal{E})$ symmetries, the resulting system possesses $\prod_{r = 1}^2 T^{r}$ (originating from $\prod_{r = 1}^2 S^{(r) }D^{o,e} D^{e,e}$ in the original model) and $\prod_{r = 1}^2 \eta^{\mathsf{CZ},r}$ (originating from $\prod_{r = 1}^2 S^{(r)}$ in the original model) symmetries. Here, $\eta^{\mathsf{CZ},r} = \prod_j (I + Z^{(r)}_j + Z^{(r)}_{j+1} - Z^{(r)}_j Z^{(r)}_{j+1})/2$ defines a non-onsite $\mathbb{Z}_2$ symmetry. We note that the resulting system only involves invertible symmetries and exhibits a type-III anomaly \cite{wang2015bosonic}. Similar to the first example, the resulting local order parameters $\mathcal{O}''_j$ and $\mathcal{Q}''_j$ can be easily identified by applying the corresponding transformations to $\mathcal{O}_j$ and $\mathcal{Q}_j$, respectively. One can also easily find that $\mathcal{O}''_j$ is charged under $\prod_{r = 1}^{2} T^{(r)}$ and $\prod_{r = 1}^{2} T^{(r)}\eta^{\mathsf{CZ},r}$ while transforming trivially under $\prod_{r = 1}^{2} \eta^{\mathsf{CZ},r}$. Conversely, $\mathcal{Q}''_j$ is charged under $\prod_{r = 1}^{2} \eta^{\mathsf{CZ},r}$ and $T^{(r)}$, while transforming trivially under $\prod_{r = 1}^{2} T^{(r)}\eta^{\mathsf{CZ},r}$. Therefore, the original non-invertible DQCP has now become an invertible DQCP between the $(\prod_{r = 1}^{2} T^{(r)}, \prod_{r = 1}^{2} T^{(r)}\eta^{\mathsf{CZ},r})$-breaking and $(\prod_{r = 1}^{2} \eta^{\mathsf{CZ},r}, \prod_{r = 1}^{2} T^{(r)})$-breaking phases after further applying the twisted gauging.
We summarize the properties of this invertible DQCP—including the symmetry generators (aside from the spin-flip and exchange symmetries), the ground-state degeneracies (GSDs) of each phase, the fixed-point wavefunctions, and the symmetries preserved by each phase—in Fig.~\ref{fig:newDQCP}(d).

\section{Conclusion and Outlook}
\label{sec:outlook}

In this work, we identified a single-parameter tuned transition between distinct non-invertible spontaneously symmetry breaking phases in coupled one-dimensional spin chains. In addition to exhibiting hallmarks of DQCP such such as self-duality, these critical points also host phenomena where non-invertible symmetries play an important role. For example, the critical point hosts a continuous emergent non-invertible symmetry. They are also robust under gauging spin-flip symmetries, which allows the construction of new families of related DQCPs.

Several natural directions emerge from our results. First, our construction requires two coupled $\mathbb{Z}^{o}_2 \times \mathbb{Z}^e_2$ spin chains, along with additional symmetry generators: $S^{(1)} S^{(2)}$, the swap between even and odd sites; $\mathcal{E}$, the exchange symmetry between the two copies; and $D^{o,1} D^{e,1} D^{o,2} D^{e,2}$, the Kramers--Wannier transformation acting on all four spin chains. Although the symmetries we impose may appear artificial, they can arise naturally in the ``double-state'' formulation of topological orders under local decoherence~\cite{bao2023mixed,chen2024unconventional}. Nonetheless, identifying a simpler microscopic model that exhibits a DQCP with a smaller set of symmetries generators would be desirable. Such a minimal realization could also facilitate numerical studies or experimental realizations.

Second, there exists an emergent continuous non-invertible ``cosine” symmetry in the vicinity of the critical surface associated with the DQCPs (see Sec.\ref{sec:alternative}). It would be interesting to investigate whether this symmetry, combined with the symmetry enforced in Eq.~\eqref{Eq:twocopy_lowenergy}, forbids the existence of a symmetric gapped phase, and thus provides a more transparent explanation of the DQCP we have identified.
{Another interesting aspect of the possibility of an emergent anomaly is that we can utilize the robustness of the non-invertible DQCP under gauging spin-flip symmetries to map our system to DQCPs that exhibit an \textit{intrinsic} anomaly. As explicitly shown in Sec.~\ref{sec:gauging}, we can map the non-invertible DQCP (with no intrinsic anomaly) we identified to invertible DQCPs that exhibit either an intrinsic LSM anomaly [Fig.~\ref{fig:newDQCP}(c)] or an intrinsic type-III anomaly [Fig.~\ref{fig:newDQCP}(d)]. This leads us to conjecture that an intrinsic anomaly in the gauged system can manifest as an emergent anomaly in the original (i.e., ungauged) system.
}

Finally, while our work provides a concrete realization of DQCPs between non-invertible SSB phases in one spatial dimension, it is natural to ask whether analogous transitions can exist in higher dimensions.

\acknowledgments We thank Arkya Chatterjee, John McGreevy, Shu-Heng Shao, and Zhehao Zhang for helpful discussions. Part of this work was completed during the KITP workshop ``Generalized Symmetries in Quantum Field Theory: High Energy Physics, Condensed Matter, and Quantum Gravity'' and we thank the workshop's organizers and participants. This research was supported in part by grant NSF PHY-2309135 to the Kavli Institute for Theoretical Physics (KITP).

\appendix
\section{Identifying Ground-State Degeneracy using the Baxter transformation}
\label{sec:Baxter}

As discussed in Sec \ref{sec:baxter_short}, the Baxter transformation $B$ maps the system in the $\mathbb{Z}_2$-charged sector $(u^o, u^e)$ and $\mathbb{Z}_2$-twisted sector $(t^o, t^e)$  to the $\mathbb{Z}_2$-charged sector $(u^X,u^Z ) = (u^o, u^e)$ and the $\mathbb{Z}_2$-twisted sector $(t^X, t^Z) = (t^o+u^e, t^e+u^o)$, respectively. 
Note that while $B$ is non-invertible over the entire Hilbert space, it acts as a unitary 
operator in each given $\mathbb{Z}_2$-charged sector $(u^o, u^e)$ and $\mathbb{Z}_2$-twisted 
sector $(t^o, t^e)$.
These properties is essential to deduce the total ground-state degeneracy (GSD) $d^{(t^o, t^e)}_{{\text{total}}}$ in any twisted sector $(t^o, t^e)$ of the original model based on those of the gauged system, which we now illustrate the basic idea. We also note that a similar technique has been used in the context of using Kennedy-Tasaki transformation to study various gapless symmetry protected topological phases \cite{li2024noninvertible, li2023intrinsically}.

Denoting the ground-state energy in the original and the gauged model as $E_{(u^o,u^e)}^{(t^o, t^e)}$ and $\tilde{E}_{(u^X,u^Z)}^{(t^X, t^Z)}$ respectively, Eqs.\eqref{Eq:uoe_uxz}  and \eqref{Eq:toe_uxz} imply
\begin{equation}
\label{Eq:sectors_orignal_dual}
E_{(u^o,u^e)}^{(t^o, t^e)} = \tilde{E}_{(u^o, u^e)}^{(t^o+u^e, t^e+u^o )} .
\end{equation}
To find the ground-state degeneracy of the original model in a given twisted sector $(t^o, t^e)$, we will first compare the ground-state energies in the four charge sectors $(u^o, u^e) = (0/1, 0/1)$, which, according to Eq.\eqref{Eq:sectors_orignal_dual}, are equivalent to the gauged model in certain symmetry and twist sectors. One can then identify the lowest-energy sector and the sectors whose ground-state energy difference from the lowest one is exponentially suppressed by the system size. We denote the set of those sectors, including the lowest-energy sector, as $\mathcal{G}^{(t^o, t^e)}$. 
Next, we will find the degeneracy $d^{(t^o, t^e)}_{(u^o, u^e)} $ for all $(u^o, u^e) \in \mathcal{G}^{(t^o, t^e)}$, which is again equivalent to degeneracy of the gauged model $\tilde{d}^{(t^X, t^Z) = (t^o+ u^e, t^e + u^o)}_{(u^X, u^Z) = (u^o, u^e)}$. The ground-state degeneracy in the twisted sector $(t^o, t^e)$ can then be obtained as
\begin{equation}
\label{Eq:d_total}
    d^{(t^o, t^e)}_{\text{total}} = \sum_{(u^o, u^e) \in \mathcal{G}} d^{(t^o, t^e)}_{(u^o, u^e)}.
\end{equation}
Since we will mostly focus on the original system with periodic boundary conditions (i.e., $t^o = t^e = 0$), we will remove the superscript when considering this sector for notational simplicity. Therefore, Eq.~\eqref{Eq:sectors_orignal_dual} and Eq.\eqref{Eq:d_total} in this sector can then be written as
\begin{equation}
\label{Eq:gs_notwist}
\begin{aligned}
E_{(u^o,u^e)}^{} & = \tilde{E}_{(u^o, u^e)}^{(u^e, u^o )},
\end{aligned}
\end{equation}
and 
\begin{equation}
\label{Eq:d_notwist}
d^{}_{\text{total}} = \sum_{(u^o, u^e) \in \mathcal{G}} d^{}_{(u^o, u^e)},
\end{equation}
respectively.
We note that if the original model can be written as a sum of commuting projectors, the ground-state degeneracy can often be determined easily without relying on the above process. However, as we will demonstrate in the following, Eq.~\eqref{Eq:gs_notwist} provides 
a way to determine the ground-state degeneracy based solely on the properties of the phase in the gauged model, without requiring the explicit form of the Hamiltonian. Furthermore, most of the interesting phases we identified in Sec.\ref{sec:identify_phases} cannot be expressed as a sum of local commuting projectors, 
making Eq.~\eqref{Eq:gs_notwist} particularly useful.

\subsection{Ground-state degeneracy of $(D^oD^e, S D^o D^e)$-breaking phase}.
\\
We now determine the ground-state degeneracy in the $(D^o D^e, S D^o D^e)$-breaking phase using insights from the VBS phase in the gauged model (see Table.\ref{table:2}), along with the topological properties of the Baxter transformation discussed in Sec.\ref{sec:Baxter}. Recall the basic idea is that we will first use Eq.\eqref{Eq:gs_notwist}, i.e., $E_{(u^o,u^e)}^{} = \tilde{E}_{u^o, u^e)}^{(u^e, u^o )}$ with $E (\tilde{E})$ the ground-state energy in the original(dual) model, to identify the lowest-energy sector and the sectors whose ground-state energy difference from the lowest one is exponentially suppressed by the system size. Denoting the set of those sectors as $\mathcal{G}$, the GSD can be computed using Eq.\eqref{Eq:d_notwist}, i.e. $d^{}_{\text{total}} = \sum_{(u^o, u^e) \in \mathcal{G}} d^{}_{(u^o, u^e)}$. 

To begin with, we note that the charge-neutral sector $u^o = u^e=0$ has the lowest energy since it corresponds to the sector $[(u^X,u^Z  ), (t^X, t^Z)] = [(0,0 ), (0,0)]$, which has the lowest energy in the gauged model.
Interestingly, the ground-state degeneracy in this sector is $d_{(u^o , u^e )=(0,0 ) } = 2$. To see this, we note that the gauged model is the VBS phase that respects both $\eta^X$ and $\eta^Z$ symmetry but breaks the translation $T$. This implies in the sector $[(u^X, u^Z  ), (t^X, t^Z)] = [(0,0 ), (0,0)]$, there are two ground states related to each other by $T$, resulting in the two-fold degeneracy in this sector.
Now, consider the $(u^o, u^e) = (0,1)$ sector, corresponding to $[(u^X, u^Z  ), (t^X, t^Z)]= [(0,1 ), (1,0)]$. Since the gauged model respects the $\eta^X$ symmetry, the cost of the $\mathbb{Z}_2^X$-twist is exponentially suppressed by the system size, and thus $ \tilde{E}_{(0,1 )}^{(1,0)} \approx \tilde{E}_{(0,0 )}^{(0,0)}$ (equivalently, ${E}_{(0,1)} \approx {E}_{(0,0 )}$). However, we now have $d_{(0,1 )}=1$ since the ground state in the sector $[(u^X, u^Z  ), (t^X, t^Z)]= [(0,1 ), (1,0)]$ is unique. This arises from the anomalous nature of the gauged system: in the presence of $\mathbb{Z}_2^X$ twist, the translational symmetry is modified as $T \rightarrow X_L T $, which exchanges the sector $[(0,0 ), (1,0)]$ and $[(0,1 ), (1,0)]$. Since the gauged model spontaneously breaks the translation, each of the sector has only one lowest energy state.
Finally, consider the sectors $(u^o, u^e) = (1,0)$ and $(1,1)$. Using the same logic as the $(u^o, u^e) = (0,1)$ sector, one finds that ${E}_{(0,1 )} \approx {E}_{(1,1)}  \approx {E}_{(0,0)}$ and $d_{(1,0)} = d_{(1,1)} = 1$. 
To summarize, we find $\mathcal{G} = \{ (0,0), (0,1), (1,0), (1,1)\}$ and $d_{(0,0)}=2,\  d_{(0,1)}=d_{(1,0)}=d_{(1,1)}=1$. Therefore,  the total ground-state degeneracy [computed using Eq.\eqref{Eq:d_notwist}] is $d_{\text{total}} = \sum_{(u^o, u^e) \in \mathcal{G}} d_{(u^o, u^e)} = 5$. 

\subsection{Ground-state degeneracy of the symmetric phase and its topological response}.

\paragraph{Ground state degeneracy}.
\\
We now show that the symmetric phase we identified in Sec.\ref{sec:identify_phases} indeed has a unique ground state using insights from the $(\eta^Z, \eta^X)$-breaking phase in the gauged model (see Table.\ref{table:2}).
First, the charge-neutral sector $u^o = u^e=0$ has the lowest energy and $d_{(0,0)} = 1$ since it corresponds to the sector $[(u^X, u^Z  ), (t^X, t^Z)] = [(0,0 ), (0,0)]$, which has a unique lowest-energy state in the gauged model. Now consider $(u^o, u^e) = (1, u^e)$, corresponding to $[(u^X,  u^Z), (t^X, t^Z)]= [(1,u^e ), (u^e,1)]$.
Since the gauged model breaks the $\eta^Z$ symmetry, introducing a $\mathbb{Z}_2^Z$ twist (i.e. $t_Z = 1$)  costs finite energy. 
Therefore, $\tilde{E}_{(1,u^e )}^{ (u^e,1)} > \tilde{E}_{(0,0)}^{(0, 0)} $ (and thus $E_{(1,u^e)} > E_{(0,0)} $) irrespective of $u^e$. 
Similarly $\tilde{E}_{(u^o,1)}^{ (1,u^o)} > \tilde{E}_{(0,0)}^{(0, 0)} $ (and thus $E_{(u^o,1)} > E_{(0,0)} $) irrespective of $u^o$ as the gauged model breaks the $\eta^X$ symmetry.
To summarize, we find $\mathcal{G} = \{ (0,0)\}$ and $d_{(0,0)}=1 $, and thus the ground state of the original model [computed using Eq.\eqref{Eq:d_total}] is unique.

\paragraph{Topological response}.
\\
One can further show that the symmetric phase is an SPT state with the property that threading a $\mathbb{Z}^o_2$ flux pumps the $\mathbb{Z}^e_2$ charge, which is a hallmark of $\mathbb{Z}^o_2 \times \mathbb{Z}^e_2 $ SPT state \cite{chen2014symmetry}.
We now demonstrate this using the topological property of Baxter transformation in the twisted sectors mentioned in Eq.\eqref{Eq:sectors_orignal_dual}, i.e., $E_{(u^o,u^e)}^{(t^o, t^e)} = \tilde{E}_{(u^o, u^e)}^{(t^o+ u^e, t^e+u^o )} $. 
Since the gauged model breaks both $\eta^Z$ and $\eta^X$ symmetry, the lowest-energy sectors in the gauged model are forced to have no twist, i.e., $(t^X, t^Z) = (0,0)$. Using $E_{(u^o,u^e)}^{(t^o, t^e)} = \tilde{E}_{(u^o, u^e)}^{(t^o+ u^e, t^e+u^o )} $, this implies  $\mathcal{G}^{(t^o, t^e)} = \{  (u^o, u^e) = (t^e, t^o) \}$. Besides, the GSD in each $(u^X, u^Z) = (u^o, u^e)$ sector is unique as the $(\eta^Z,\eta^X)$-breaking phase respects translational symmetry. Therefore, the ground state is $\mathbb{Z}^e_2$ charged under a $\mathbb{Z}^o_2$ twist and vice versa.
{Finally, we remark that although the symmetric phase resembles the cluster state with respect to the $\eta^o$ and $\eta^e$ symmetries, it is not adiabatically connected to it in the presence of the swap symmetry, since the cluster state does not respect this symmetry. 
}
%\bibliography{bibs}

\end{document}